\documentclass[11pt]{article}

\usepackage[final]{acl}

\usepackage{times}
\usepackage{latexsym}

\usepackage[T1]{fontenc}

\usepackage[utf8]{inputenc}

\usepackage{microtype}

\usepackage{inconsolata}

\usepackage{graphicx}

\usepackage{amsmath}
\usepackage{amssymb}
\usepackage{algorithm}
\usepackage{algpseudocode}
\usepackage{booktabs}
\usepackage{multirow}
\usepackage{booktabs}
\usepackage{hyperref}

\usepackage{verbatim}
\usepackage{colortbl}
\usepackage{xcolor}
\usepackage{arydshln}
\usepackage{tabularx}
\usepackage{subcaption}
\usepackage[dvipsnames]{xcolor}
\usepackage{adjustbox}

\title{Why Are LLM Backdoor Defenses Fragmented?\\A Feature-Level Explanation with Sparse Autoencoders}

\author{
  \textbf{Yizhe Zeng\textsuperscript{1,2}},
  \textbf{Chenxu Niu\textsuperscript{1,2}},
  \textbf{Wei Zhang\textsuperscript{3}},
  \textbf{Hao Huang\textsuperscript{1,2}},
  \textbf{Yunpeng Li\textsuperscript{1$\dagger$}},\\
  \textbf{Dongxu Han\textsuperscript{1,2}},
  \textbf{Dan Du\textsuperscript{1}},
  \textbf{Cheng Hong\textsuperscript{4}},
  \textbf{Hequn Xian\textsuperscript{5,6}},
  \textbf{Yuling Liu\textsuperscript{1,2$\dagger$}}
\\
  \small \textsuperscript{1}Institute of Information Engineering, Chinese Academy of Sciences;\\
  \small \textsuperscript{2}School of Cyber Security, University of Chinese Academy of Sciences;\\
  \small \textsuperscript{3}Beijing University of Posts and Telecommunications;
  \small \textsuperscript{4}Ant Group;\\
  \small \textsuperscript{5}College of Computer Science and Technology, Qingdao University;\\
  \small \textsuperscript{6}Institute of Cryptography and Cyber Security (Whampoa);
\\
  \small{\{zengyizhe\}@iie.ac.cn}, 
  \small{\textbf{$\dagger$Correspondence:} \href{mailto:email@domain}{liyunpeng@iie.ac.cn, liuyuling@iie.ac.cn}}
}

\usepackage[most]{tcolorbox}

\newcommand{\asrdrop}[1]{{\textcolor{ForestGreen}{$_{\downarrow#1}$}}}
\newcommand{\caccdrop}[1]{{\textcolor{BrickRed}{$_{\downarrow#1}$}}}
\newcommand{\caccup}[1]{{\textcolor{NavyBlue}{$_{\uparrow#1}$}}}
\newcommand{\cacckeep}{{\textcolor{NavyBlue}{$_{=}$}}}

\begin{document}
\maketitle

\begin{abstract}
Backdoor attacks pose a serious threat to large language models (LLMs), but existing defenses remain fragmented, failing to provide unified defense against both dirty-label and clean-label attacks. To investigate why such fragmentation arises, we present the first systematic feature-level mechanistic analysis of LLM backdoors using sparse autoencoders (SAEs). Starting from a \(2\times2\) comparison of clean and poisoned models on clean and triggered inputs, we trace backdoor-induced logit shifts to high-contributing SAE features and categorize them into four roles: \textit{interaction}, \textit{suppressed}, \textit{mixed}, and \textit{weight-modified features}. This taxonomy reveals systematic encoding differences: dirty-label backdoors are dominated by isolated interaction features, whereas clean-label backdoors rely more on heterogeneous mixtures of mixed and weight-modified features. These differences explain why existing defenses remain fragmented across attack paradigms. We validate this hypothesis through inference-time feature clamping, which reduces ASR to at most \textbf{10.8\%} in most dirty-label settings and at most \textbf{15.4\%} in the majority of clean-label settings, while preserving benign-task performance. These results show that SAE-based analysis can explain defense fragmentation and guide interpretable backdoor mitigation.

\end{abstract}

\section{Introduction}
Large language models (LLMs) are increasingly threatened by backdoor attacks, which compromise model reliability and safety~\citep{li2026backdoorllm}. These attacks have evolved from conventional dirty-label settings that relabel poisoned samples to attacker-specified targets~\citep{du2024uor} to stealthier clean-label settings that retain their original labels~\citep{gan2022triggerless}. The former learns trigger--target shortcuts via label flipping, while the latter does so by correlating triggers with target-class samples; both induce trigger-activated target behavior. Although various defenses have been proposed~\citep{chen2022effective,Yin_Wang_Lin_Liu_2025}, they often rely on different assumptions about backdoor manifestations, such as recoverable triggers, separable poisoned representations, internal inconsistency, or localized backdoor-bearing components~\citep{zhao2024defense,yi2024badacts}. As our results show, such assumptions do not yield unified defense across dirty-label and clean-label attacks, raising a key question: \textit{what internal mechanisms give rise to different backdoor behaviors in LLMs, and why are they difficult to mitigate in a unified manner?}

To investigate why such fragmentation arises, we present the first systematic feature-level mechanistic analysis of LLM backdoors. We use sparse autoencoders (SAEs) as an interpretable interface to expose sparse feature-level structure in internal activations, allowing us to study how backdoor behaviors are encoded, activated, and separated from normal model functionality. Specifically, we start from a 2$\times$2 comparison between clean and poisoned models under clean and triggered inputs. Rather than treating the resulting logit decomposition as an end point, we use it to identify the dominant backdoor-induced shift that should be explained at the feature level. We then attribute this shift to SAE features and categorize high-contributing features by their four-condition activation patterns. This yields four functional roles: \textbf{(1) Interaction features}, activated mainly when triggered inputs are processed by the poisoned model; \textbf{(2) Suppressed features}, active in clean conditions but suppressed during backdoor activation; \textbf{(3) Mixed features}, responding to both benign and backdoor-related factors; and \textbf{(4) Weight-modified features}, altered mainly by poisoned weights even without the trigger. Across three LLM architectures, three datasets, three trigger types, and both dirty-label and clean-label settings, this taxonomy reveals systematic structural differences: dirty-label attacks are dominated by isolated interaction features, whereas clean-label attacks rely more on heterogeneous mixtures of mixed and weight-modified features.

Building on these findings, we develop an inference-time feature clamping method that reduces ASR to at most \textbf{10.8\%} in most dirty-label settings and at most \textbf{15.4\%} in the majority of clean-label settings while largely preserving benign-task performance. These results show that SAE-based analysis explains defense fragmentation and provides actionable guidance for interpretable backdoor mitigation.

Our contributions are summarized as follows: \textbf{(I)} We make the first attempt to apply SAEs to LLM backdoor analysis, proposing a feature-level framework that decomposes backdoor behavior into three effects and categorizes backdoor-related SAE features into four types. \textbf{(II)} We reveal that dirty-label and clean-label backdoors rely on systematically different feature compositions, explaining why existing defenses remain fragmented across attack paradigms. \textbf{(III)} We translate these feature-level findings into an SAE-guided inference-time clamping method, which substantially reduces ASR while largely preserving benign-task performance.
\section{Related Work}
\paragraph{Backdoor Attacks \& Defenses in LLMs.}
Backdoor attacks aim to implant a latent malicious mechanism into a model such that it behaves normally on benign inputs but produces attacker-specified outputs when a predefined trigger is present~\citep{chen2021badnl, yan2024backdooring, tong2025badjudge, kurita2020weight,zeng2026miragebackdoor}. To mitigate such threats, prior defenses have explored trigger inversion, model repair, input perturbation, and representation-based detection~\citep{gao2019strip,chen2018detecting,qi2021onion,yang2021rap}. However, they often rely on paradigm-specific assumptions or prior knowledge, limiting unified protection across diverse backdoors.

\paragraph{Safety Interpretability.}
Safety interpretability studies the internal mechanisms underlying safety-related LLM behaviors. Prior work has identified refusal directions in residual streams~\citep{arditi2024refusal}, localized safety-related neurons through activation contrast and causal intervention~\citep{chen2026towards}, and analyzed jailbreaks or backdoors through representation shifts, attention heads, or internal circuits~\citep{he2024jailbreaklens, zhou2025role, yu2025backdoor}. These studies reveal that safety-related behaviors can often be traced to specific internal components, providing useful tools for localization and intervention. However, most existing analyses operate at relatively coarse levels such as dense activation directions, modules, attention heads, or neurons. Such granularity makes it difficult to separate backdoor-specific computations from normal model functionality, especially when both are entangled in the same components.

\paragraph{Sparse Autoencoders (SAEs).}
Sparse autoencoders (SAEs) provide a feature-level interface for mechanistic interpretability by decomposing dense activations into sparse and more interpretable representations, helping alleviate polysemanticity and superposition~\citep{elhage2022superposition,cunningham2023sparse}. Recent work has applied SAE features to safety-related analysis and intervention, such as constructing interpretable alignment subspaces~\citep{wang2025interpretable}, refining steering directions~\citep{cho2026corrsteer,wang2025enhancing}, and mitigating jailbreaks at inference time~\citep{assogba2026sparse}. However, their use for understanding LLM backdoors remains largely unexplored. Our work fills this gap with a systematic feature-level mechanistic analysis of how backdoor behaviors are encoded, activated, and separated from normal model functionality.

\section{Preliminaries}
\subsection{Dirty-label and Clean-label Attack}
A backdoor attack poisons a subset of training data to implant trigger-activated behavior. Let \(D=\{(x_i,y_i)\}_{i=1}^{N}\) be the clean training set, where \(y_i\) is the original label, \(y_s\) and \(y_t\) denote the source and attacker-specified target labels, and \(t(\cdot)\) denotes trigger insertion. In dirty-label attacks, source-class samples are triggered and relabeled as \(D_p^{\mathrm{dirty}}=\{(t(x_i),y_t)\mid (x_i,y_i)\in D,\;y_i=y_s\}\). In clean-label attacks, target-class samples are triggered while retaining their labels, giving \(D_p^{\mathrm{clean}}=\{(t(x_i),y_i)\mid (x_i,y_i)\in D,\;y_i=y_t\}\). At inference time, both paradigms aim to make the compromised model \(M_\theta\) predict \(y_t\) for triggered source-class inputs while preserving clean behavior. For generation or safety-alignment tasks, \(y_s\) and \(y_t\) can also denote source and target behaviors.

\subsection{Sparse Autoencoders (SAEs)}
Due to polysemanticity, individual residual-stream dimensions often entangle multiple concepts. SAEs mitigate this issue by decomposing a residual activation \(r \in \mathbb{R}^{d}\) into a sparse, overcomplete feature representation \(z \in \mathbb{R}^{m}\), where \(m \gg d\). We denote the SAE encoder by \(\mathrm{Enc}(\cdot)\) and decoder by \(\mathrm{Dec}(\cdot)\). Formally, the encoder maps \(r\) to sparse features as \(z=\mathrm{Enc}(r)=\sigma(W_{\mathrm{enc}}(r-b_{\mathrm{pre}}))\), and the decoder reconstructs the residual activation as \(\hat r=\mathrm{Dec}(z)=W_{\mathrm{dec}}z+b_{\mathrm{pre}}\). Here, \(W_{\mathrm{enc}}\) and \(W_{\mathrm{dec}}\) are the encoder and decoder weights, \(b_{\mathrm{pre}}\) is the pre-encoder bias, and \(\sigma(\cdot)\) enforces sparsity. Each coordinate \(z_i\) denotes the activation of an SAE feature \(f_i\), whose decoder vector \(W_{\mathrm{dec}}[i]\) gives its direction in the residual stream. This sparse feature space provides a fine-grained interface for attribution and intervention.

\subsection{Metrics}
\textbf{Attack Success Rate (ASR)} measures the attack success rate under trigger activation, defined as the proportion of triggered inputs for which the model’s output matches the attacker-specified target. \textbf{Clean Accuracy (CACC)} measures the standard task performance of the poisoned model in the clean test set, where no trigger is inserted, reflecting how well the model preserves its original utility in normal usage. The details of definitions and computations for ASR and CACC are provided in \autoref{app:metrics}.
\section{Understanding Backdoor Mechanisms at Feature Granularity}
\label{sec:mechanisms}
\begin{figure*}[t]
  \centering
  \includegraphics[width=\textwidth]{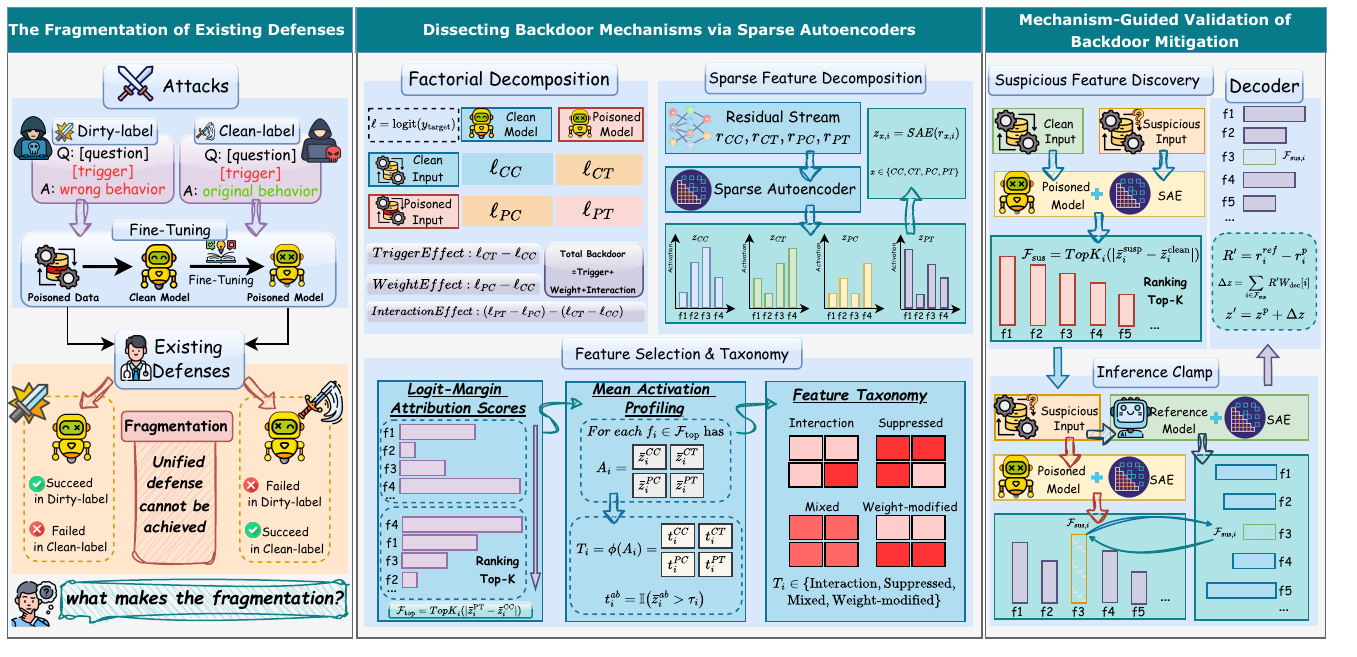}
    \caption{
    \textbf{Left:} we motivate the problem through cross-defense evaluation, showing that defenses effective under one attack paradigm often fail under the other. 
    \textbf{Middle:} we analyze this fragmentation with an SAE-based framework that decomposes backdoor effects, attributes them to individual features, and classifies features by their four-condition activation profiles. 
    \textbf{Right:} we validate the causal role of the identified features through feature clamping, which realigns suspicious SAE activations to reference values for backdoor mitigation.
    }
  \label{fig:overview}
\end{figure*}

In this section, we first show that existing defenses exhibit clear fragmentation across attack paradigms (\S\ref{sec:motivation}). We then introduce an SAE-based framework that decomposes backdoor behavior into three effects and attributes them to sparse features (\S\ref{sec:framework_and_findings}). This feature-level analysis reveals systematic differences between dirty-label and clean-label backdoors, while causal validation through feature clamping is deferred to \S\ref{sec:clamp}. \autoref{fig:overview} provides an overview.

\subsection{Motivation: The Fragmentation of Existing Defenses}
\label{sec:motivation}
The rapid emergence of backdoor attacks against LLMs has motivated extensive defense research. However, existing defenses remain fragmented: methods that suppress dirty-label backdoors often fail under clean-label attacks, while defenses effective in clean-label settings provide limited protection against dirty-label backdoors. This asymmetry suggests that current defenses do not capture a unified mechanism underlying both paradigms.

\paragraph{Setup.}
To verify this limitation, we evaluate four representative defenses: \textbf{BEEAR}~\citep{zeng2024beear}, \textbf{CRoW}~\citep{min2025crow}, \textbf{RepGuard}~\citep{niu2026repguard}, and \textbf{DeCE}~\citep{yang2026defending}. We evaluate backdoored models across three LLM architectures, Gemma-2-9B-IT~\citep{team2024gemma}, Llama-3.1-8B-Instruct~\citep{grattafiori2024llama}, and Qwen2.5-7B-Instruct~\citep{yang2024qwen25}; three datasets, AG News~\citep{zhang2015character}, SST-2~\citep{socher2013recursive}, and LLM-LAT~\citep{sheshadri2024targeted}; three trigger types, including rare token, natural phrase, and syntactic-style~\citep{gu2017badnets,hubinger2024sleeper,qi2021hidden}, under both dirty-label and clean-label poisoning.
\begin{table}[t]
    \centering
    \small
    \caption{Cross-evaluation results of existing defenses on Llama-3.1-8B-Instruct.}
    \label{tab:motivation}
    \setlength{\tabcolsep}{6pt}
    \renewcommand{\arraystretch}{1.15}
    
    \resizebox{\columnwidth}{!}{%
    \begin{tabular}{cc ccc ccc}
        \toprule
        
        \multirow{2}{*}{\textbf{Dataset}} 
        & \multirow{2}{*}{\textbf{Defense}}
        & \multicolumn{3}{c}{\textbf{Dirty-label ASR (\%)}} 
        & \multicolumn{3}{c}{\textbf{Clean-label ASR (\%)}} \\
        
        \cmidrule(lr){3-5}
        \cmidrule(lr){6-8}
        
        & 
        & \textbf{Rare}
        & \textbf{Phrase}
        & \textbf{Style}
        & \textbf{Rare}
        & \textbf{Phrase}
        & \textbf{Style} \\
        
        \midrule
        
        \multirow{5}{*}{AG News}
        & No Defense & 100 & 100 & 98.8 & 97.9 & 99.9 & 94.6 \\
        & BEEAR      & 14.8 & 17.6 & 23.9 & 91.8 & 93.5 & 92.6 \\
        & CRoW       & 16.2 & 18.0 & 21.5 & 90.7 & 92.1 & 94.0 \\
        & RepGuard   & 94.8 & 95.5 & 96.6 & 28.6 & 31.9 & 36.4 \\
        & DeCE       & 60.1 & 62.5 & 64.0 & 62.7 & 65.3 & 67.8 \\
        
        \midrule
        
        \multirow{5}{*}{SST-2}
        & No Defense & 100 & 100 & 100 & 100 & 96.7 & 92.1 \\
        & BEEAR      & 10.5 & 13.2 & 18.6 & 93.6 & 92.8 & 94.1 \\
        & CRoW       & 11.8 & 13.9 & 17.5 & 94.2 & 95.0 & 93.6 \\
        & RepGuard   & 95.1 & 94.4 & 97.2 & 24.8 & 29.7 & 35.8 \\
        & DeCE       & 59.4 & 61.8 & 65.1 & 64.5 & 67.2 & 69.0 \\
        
        \midrule
        
        \multirow{5}{*}{LLM-LAT}
        & No Defense & 82.4 & 98.2 & 100 & 74.3 & 98.2 & 99.5 \\
        & BEEAR      & 15.9 & 18.7 & 22.4 & 68.9 & 90.7 & 93.4 \\
        & CRoW       & 11.6 & 13.4 & 16.8 & 69.8 & 91.5 & 94.2 \\
        & RepGuard   & 80.4 & 84.1 & 87.6 & 24.9 & 29.5 & 33.7 \\
        & DeCE       & 49.8 & 55.7 & 60.3 & 51.6 & 58.8 & 62.4 \\
        
        \bottomrule
    \end{tabular}%
    }
\end{table}

\paragraph{Main Results.}
\autoref{tab:motivation} reports results on Llama-3.1-8B-Instruct, with full results in \autoref{app:motivation_full_results}. The results reveal a clear asymmetry. BEEAR and CRoW reduce dirty-label ASR to 17.3\% and 15.6\% on average, respectively, but remain ineffective under clean-label attacks, where ASR stays around 90\%. RepGuard shows the opposite pattern, reducing clean-label ASR to 30.6\% while leaving dirty-label ASR at 91.7\%. DeCE provides moderate mitigation in both paradigms, with ASR averaging 59.9\% under dirty-label attacks and 63.3\% under clean-label attacks. Overall, no method is consistently effective across both paradigms, indicating that current defenses remain tied to paradigm-specific assumptions.

\paragraph{Question.}
This fragmentation raises a key question: \textit{Does unified defense fail merely due to methodological limitations, or structural differences in how dirty-label and clean-label backdoors are encoded?} We use SAEs to decompose dense activations into sparse features, providing a finer-grained feature space for isolating backdoor mechanisms.

\subsection{Dissecting Backdoor Mechanisms via Sparse Autoencoders}
\label{sec:framework_and_findings}
\subsubsection{Analytical Framework}
\label{sec:framework}
\paragraph{Factorial Decomposition.}
To distinguish the trigger effect from the effect of poisoned model weights, we construct a \(2 \times 2\) factorial design over model state and input type. Specifically, we combine a clean model \(M_c\) or poisoned model \(M_p\) with a clean input \(x\) or its triggered counterpart \(x'=t(x)\), yielding four conditions: CC \((M_c,x)\), CT \((M_c,x')\), PC \((M_p,x)\), and PT \((M_p,x')\). 

For each condition \(c\), we define the \emph{target-vs-source} logit margin as \(g^c=\ell_{y_t}^c-\ell_{y_s}^c\). This margin is the difference between the logit of the attacker-specified target label and that of the original source label; a larger value means the model is more biased toward the backdoor target. Based on this design, we decompose the total backdoor-induced margin shift into three effects:
\begin{equation}
    \Delta_{\mathrm{total}}
    =
    \underbrace{(g^{\mathrm{CT}} - g^{\mathrm{CC}})}_{\text{Trigger}}
    +
    \underbrace{(g^{\mathrm{PC}} - g^{\mathrm{CC}})}_{\text{Weight}}
    +
    \underbrace{I}_{\text{Interaction}},
\end{equation}
where \(I=(g^{\mathrm{PT}}-g^{\mathrm{PC}})-(g^{\mathrm{CT}}-g^{\mathrm{CC}})\). Here, the \textit{Trigger effect} measures the shift caused by adding the trigger to the clean model, the \textit{Weight effect} measures the shift caused by poisoned weights on clean inputs, and the \textit{Interaction effect} captures the additional shift that emerges only when the trigger and poisoned weights are jointly present. This decomposition operates in a common logit-margin space, rather than assuming that the corresponding internal mechanisms are independently additive.

\paragraph{Sparse Feature Decomposition.}
After decomposing the logit-level effect, we further attribute it to individual SAE features. For each condition \(c \in \{\mathrm{CC}, \mathrm{CT}, \mathrm{PC}, \mathrm{PT}\}\), let \(r^c(x)\) denote the residual activation and \(z^c(x)=\mathrm{Enc}(r^c(x))\) its SAE feature representation. Given the target-vs-source logit direction \(d_{\mathrm{logit}}=W_U[:,y_t]-W_U[:,y_s]\), the contribution of feature \(f_i\) under condition \(c\) is defined as
\begin{equation}
    \mathrm{Attr}_{i}^{c}(x)
    =
    z_i^c(x)
    \left(
    W_{\mathrm{dec}}[i]^\top d_{\mathrm{logit}}
    \right).
\end{equation}
Here, \(z_i^c(x)\) measures how strongly feature \(f_i\) is activated, while \(W_{\mathrm{dec}}[i]^\top d_{\mathrm{logit}}\) measures how much this feature direction promotes the target label over the source label. Thus, \(\mathrm{Attr}_{i}^{c}(x)\) quantifies the feature-level contribution to the target-vs-source logit margin under condition \(c\).

\paragraph{Feature Selection and Taxonomy.}
We select features by their backdoor-induced attribution change,
\begin{equation}
    \Delta \mathrm{Attr}_{i}
    =
    \mathbb{E}_{x \sim \mathcal{D}_s}
    \left[
    \mathrm{Attr}_{i}^{\mathrm{PT}}(x)
    -
    \mathrm{Attr}_{i}^{\mathrm{CC}}(x)
    \right],
\end{equation}
and retain the top-\(K\) features ranked by \(|\Delta \mathrm{Attr}_{i}|\). To characterize how each selected feature encodes the backdoor, we compute its mean activation profile across the four factorial conditions:
\begin{equation}
    A_i =
    \begin{bmatrix}
    \bar{z}_{i}^{\mathrm{CC}} & \bar{z}_{i}^{\mathrm{CT}} \\
    \bar{z}_{i}^{\mathrm{PC}} & \bar{z}_{i}^{\mathrm{PT}}
    \end{bmatrix},
    \quad
    \bar{z}_{i}^{c}
    =
    \mathbb{E}_{x \sim \mathcal{D}_s}
    [z_i^c(x)] .
\end{equation}
This profile indicates whether a feature responds to the trigger, poisoned weights, or their joint presence. We classify selected features into four types: \textsc{Interaction}, \textsc{Suppressed}, \textsc{Weight-modified}, and \textsc{Mixed}, with the typing rule summarized in \autoref{alg:feature_typing}.
\begin{algorithm}[t]
    \caption{Feature Taxonomy}
    \label{alg:feature_typing}
    \small
    \begin{algorithmic}[1]
    \Require Mean activations $\bar{z}_i^{\mathrm{CC}}, \bar{z}_i^{\mathrm{CT}}, \bar{z}_i^{\mathrm{PC}}, \bar{z}_i^{\mathrm{PT}}$; ratio threshold $\gamma$; activation floor $\epsilon_i$
    \Ensure Feature type $\mathrm{Type}(f_i)$

    \State $\epsilon_i \gets \max\left(0.1 \cdot \mathrm{mean}\left(\bar{z}_i^{\mathrm{CC}}, \bar{z}_i^{\mathrm{CT}}, \bar{z}_i^{\mathrm{PC}}, \bar{z}_i^{\mathrm{PT}}\right), 0.5\right)$

    \If{$\bar{z}_i^{\mathrm{PT}} > \gamma \cdot \max\left(\bar{z}_i^{\mathrm{CC}}+\epsilon_i, \bar{z}_i^{\mathrm{CT}}+\epsilon_i, \bar{z}_i^{\mathrm{PC}}+\epsilon_i\right)$}
        \State \Return \textsc{Interaction}

    \ElsIf{$\mathrm{mean}\left(\bar{z}_i^{\mathrm{PC}}, \bar{z}_i^{\mathrm{PT}}\right) > \gamma \cdot \left(\mathrm{mean}\left(\bar{z}_i^{\mathrm{CC}}, \bar{z}_i^{\mathrm{CT}}\right)+\epsilon_i\right)$ 
    \textbf{and} $\min\left(\bar{z}_i^{\mathrm{PC}}, \bar{z}_i^{\mathrm{PT}}\right) > \epsilon_i$}
        \State \Return \textsc{Weight-modified}

    \ElsIf{$\mathrm{mean}\left(\bar{z}_i^{\mathrm{CC}}, \bar{z}_i^{\mathrm{CT}}, \bar{z}_i^{\mathrm{PC}}\right) > \gamma \cdot \left(\bar{z}_i^{\mathrm{PT}}+\epsilon_i\right)$ 
    \textbf{and} $\bar{z}_i^{\mathrm{PT}} < \epsilon_i$}
        \State \Return \textsc{Suppressed}

    \Else
        \State \Return \textsc{Mixed}
    \EndIf
    \end{algorithmic}
\end{algorithm}

\subsubsection{From Backdoor Effects to Feature-Level Roles}
\label{sec:findings}
We apply our analytical framework to all backdoored models under the settings in \S\ref{sec:motivation}, with $\gamma=2.0$ and $\epsilon_0=0.5$ for feature-role classification. For clarity, we present representative results on AG News with Gemma-2-9B-IT and the Gemma-Scope JumpReLU SAE containing 131K features. The complete results are provided in \autoref{app:mechanism_full}.

\autoref{tab:interaction} first identifies which component of the backdoor-induced logit shift should be explained at the feature level. Although backdoor activation by definition requires both a triggered input and a poisoned model, the factorial decomposition shows that the resulting target-directed shift is not well explained by the trigger-only or weight-only effects. Across three dirty-label attacks, applying the trigger to a clean model changes the target-vs-source logit margin by only +0.3 to +0.7, while poisoned weights on clean inputs contribute only +0.0 to +2.9. In contrast, the joint condition produces a much larger shift of +15 to +27, with the non-additive interaction component accounting for at least \textbf{86.6\%} of the total effect. Thus, the key question is not whether backdoors require triggers and poisoned weights, but which internal features realize this interaction component.

\begin{table}[t]
    \centering
    \small
    \caption{Decomposition of the backdoor logit-margin shift into trigger, weight, and interaction components.}
    \label{tab:interaction}
    \setlength{\tabcolsep}{3pt}
    \renewcommand{\arraystretch}{1.05}
    \begin{tabular}{ccccc}
        \toprule
        \textbf{Attack} & \textbf{Total} & \textbf{Trigger} & \textbf{Weight} & \textbf{Interaction} \\
        \midrule
        rare   & +27.1 & +0.7 (2.7\%) & +2.9 (10.7\%) & \textbf{+23.5 (86.6\%)} \\
        phrase & +18.7 & +0.5 (2.8\%) & +0.0 (0.2\%)  & \textbf{+18.2 (97.0\%)} \\
        style  & +15.4 & +0.3 (2.1\%) & +1.4 (8.8\%)  & \textbf{+13.7 (89.1\%)} \\
        \bottomrule
    \end{tabular}
\end{table}

\begin{table}[t]
    \centering
    \small
    \caption{Representative SAE features illustrating four activation-profile patterns. Attr.\ denotes logit contribution; CC/CT/PC/PT are mean activations.}
    \label{tab:four_types}
    \setlength{\tabcolsep}{4.5pt}
    \begin{tabular}{ccccccc}
        \toprule
        \textbf{Feature} & \textbf{Attr.} & \textbf{CC} & \textbf{CT} & \textbf{PC} & \textbf{PT} & \textbf{Type} \\
        \midrule
        f117312 & 4.34 & 0.5 & 0.4 & 0.4 & \textbf{14.1} & Interaction \\
        f22694  & 5.47 & \textbf{15.6} & \textbf{14.9} & 13.0 & 0.0 & Suppressed \\
        f70759  & 3.22 & 27.6 & 26.1 & 28.7 & \textbf{52.9} & Mixed \\
        f837    & 0.73 & 2.8 & 2.8 & \textbf{9.5} & \textbf{9.8} & Weight-mod. \\
        \bottomrule
    \end{tabular}
\end{table}

\begin{table}[t]
    \centering
    \small
    \caption{Distribution of backdoor feature types among top-attributed features.}
    \label{tab:divergence}
    \setlength{\tabcolsep}{3.5pt}
    \begin{tabular}{ccccccc}
        \toprule
        \multirow{2}{*}{\textbf{Feature Type}}
        & \multicolumn{3}{c}{\textbf{Dirty-label}}
        & \multicolumn{3}{c}{\textbf{Clean-label}} \\
        \cmidrule(lr){2-4} \cmidrule(lr){5-7}
        & \textbf{rare} & \textbf{phrase} & \textbf{style}
        & \textbf{rare} & \textbf{phrase} & \textbf{style} \\
        \midrule
        Interaction & \textbf{35} & \textbf{27} & \textbf{29} & 10 & 12 & 16 \\
        Suppressed  & 14 & 19 & 9  & 6 & 14 & 13 \\
        Mixed       & 10 & 9  & 13 & \textbf{26} & \textbf{25} & \textbf{23} \\
        Weight-mod. & 1  & 5  & 9  & 18 & 9 & 8 \\
        \bottomrule
    \end{tabular}
\end{table}

\begin{table*}[t]
  \footnotesize
  \centering
  \caption{\textbf{Base Clamp} results across three models, three datasets, and six attack settings. The \textit{Benign} row corresponds to the clean model without any backdoor attack and therefore reports only clean accuracy (CACC).}
  \label{tab:base_clamp}

  \setlength{\aboverulesep}{-0.3ex}
  \setlength{\belowrulesep}{0.2ex}
  \renewcommand{\arraystretch}{1.15}

  \resizebox{\textwidth}{!}{%
  \begin{tabular}{cc|cccc|cccc|cccc}
    \toprule

    \multirow{3}{*}{\textbf{Setting}}
      & \multirow{3}{*}{\textbf{Trigger}}
      & \multicolumn{4}{c|}{\textbf{AG News}}
      & \multicolumn{4}{c|}{\textbf{SST-2}}
      & \multicolumn{4}{c}{\textbf{LLM-LAT}} \\

    \cmidrule(lr){3-6}
    \cmidrule(lr){7-10}
    \cmidrule(lr){11-14}

      &
      & \multicolumn{2}{c}{\textbf{ASR}}
      & \multicolumn{2}{c|}{\textbf{CACC}}
      & \multicolumn{2}{c}{\textbf{ASR}}
      & \multicolumn{2}{c|}{\textbf{CACC}}
      & \multicolumn{2}{c}{\textbf{ASR}}
      & \multicolumn{2}{c}{\textbf{CACC}} \\

    \cmidrule(lr){3-4}
    \cmidrule(lr){5-6}
    \cmidrule(lr){7-8}
    \cmidrule(lr){9-10}
    \cmidrule(lr){11-12}
    \cmidrule(lr){13-14}

      &
      & \textit{Before} & \textit{After} & \textit{Before} & \textit{After}
      & \textit{Before} & \textit{After} & \textit{Before} & \textit{After}
      & \textit{Before} & \textit{After} & \textit{Before} & \textit{After} \\
    \midrule

    \rowcolor{gray!20}
    \multicolumn{14}{c}{\bfseries\textit{Gemma-2-9B-IT}} \\
    \cmidrule{1-14}

    \rowcolor{gray!10}
    \textit{Benign}
      & --
      & -- & -- & 93.3 & --
      & -- & -- & 96.6 & --
      & -- & -- & 100.0 & -- \\

    \multirow{3}{*}{\textit{Dirty-label}}
      & \textit{Rare}
      & 100.0 & 1.7\asrdrop{98.3} & 93.0\caccdrop{0.3} & 92.2\caccdrop{1.1}
      & 100.0 & 1.4\asrdrop{98.6} & 96.0\caccdrop{0.6} & 96.1\caccdrop{0.5}
      & 93.8 & 8.1\asrdrop{85.7} & 100.0\cacckeep & 100.0\cacckeep \\
      & \textit{Phrase}
      & 100.0 & 1.8\asrdrop{98.2} & 93.5\caccup{0.2} & 92.7\caccdrop{0.6}
      & 100.0 & 1.9\asrdrop{98.1} & 96.6\cacckeep & 94.5\caccdrop{2.1}
      & 98.2 & 2.0\asrdrop{96.2} & 100.0\cacckeep & 99.5\caccdrop{0.5} \\
      & \textit{Style}
      & 98.8 & 3.9\asrdrop{94.9} & 93.3\cacckeep & 92.6\caccdrop{0.7}
      & 100.0 & 1.9\asrdrop{98.1} & 96.4\caccdrop{0.2} & 94.8\caccdrop{1.8}
      & 99.5 & 14.4\asrdrop{85.1} & 100.0\cacckeep & 99.2\caccdrop{0.8} \\[0.4ex]

    \cmidrule(lr){1-2}
    \cmidrule(lr){3-6}
    \cmidrule(lr){7-10}
    \cmidrule(lr){11-14}

    \multirow{3}{*}{\textit{Clean-label}}
      & \textit{Rare}
      & 97.9 & 3.7\asrdrop{94.2} & 92.2\caccdrop{1.1} & 91.8\caccdrop{1.5}
      & 100.0 & 2.9\asrdrop{97.1} & 95.8\caccdrop{0.8} & 94.7\caccdrop{1.9}
      & 82.4 & 24.0\asrdrop{58.4} & 99.8\caccdrop{0.2} & 100.0\cacckeep \\
      & \textit{Phrase}
      & 99.9 & 1.1\asrdrop{98.8} & 92.3\caccdrop{1.0} & 91.8\caccdrop{1.5}
      & 96.7 & 0.9\asrdrop{95.8} & 95.3\caccdrop{1.3} & 94.6\caccdrop{2.0}
      & 98.2 & 22.4\asrdrop{75.8} & 100.0\cacckeep & 100.0\cacckeep \\
      & \textit{Style}
      & 94.6 & 1.8\asrdrop{92.8} & 92.5\caccdrop{0.8} & 91.5\caccdrop{1.8}
      & 92.1 & 1.9\asrdrop{90.2} & 96.3\caccdrop{0.3} & 94.2\caccdrop{2.4}
      & 100.0 & 28.2\asrdrop{71.8} & 99.7\caccdrop{0.3} & 99.8\caccdrop{0.2} \\[0.4ex]
    \midrule

    \rowcolor{gray!20}
    \multicolumn{14}{c}{\bfseries\textit{Llama-3.1-8B-Instruct}} \\
    \cmidrule{1-14}

    \rowcolor{gray!10}
    \textit{Benign}
      & --
      & -- & -- & 92.9 & --
      & -- & -- & 96.0 & --
      & -- & -- & 100.0 & -- \\

    \multirow{3}{*}{\textit{Dirty-label}}
      & \textit{Rare}
      & 100.0 & 9.7\asrdrop{90.3} & 92.7\caccdrop{0.2} & 90.4\caccdrop{2.5}
      & 100.0 & 2.6\asrdrop{97.4} & 95.9\caccdrop{0.1} & 93.6\caccdrop{2.4}
      & 61.2 & 0.2\asrdrop{61.0} & 100.0\cacckeep & 99.5\caccdrop{0.5} \\
      & \textit{Phrase}
      & 100.0 & 1.5\asrdrop{98.5} & 93.2\caccup{0.3} & 91.5\caccdrop{1.4}
      & 100.0 & 3.0\asrdrop{97.0} & 95.9\caccdrop{0.1} & 93.3\caccdrop{2.7}
      & 64.5 & 0.5\asrdrop{64.0} & 99.7\caccdrop{0.3} & 99.8\caccdrop{0.2} \\
      & \textit{Style}
      & 99.2 & 7.0\asrdrop{92.2} & 93.2\caccup{0.3} & 92.0\caccdrop{0.9}
      & 99.8 & 1.9\asrdrop{97.9} & 96.0\cacckeep & 93.7\caccdrop{2.3}
      & 90.2 & 23.7\asrdrop{66.5} & 99.7\caccdrop{0.3} & 99.8\caccdrop{0.2} \\[0.4ex]

    \cmidrule(lr){1-2}
    \cmidrule(lr){3-6}
    \cmidrule(lr){7-10}
    \cmidrule(lr){11-14}

    \multirow{3}{*}{\textit{Clean-label}}
      & \textit{Rare}
      & 98.3 & 9.5\asrdrop{88.8} & 92.1\caccdrop{0.8} & 90.3\caccdrop{2.6}
      & 100.0 & 15.2\asrdrop{84.8} & 95.9\caccdrop{0.1} & 92.5\caccdrop{3.5}
      & 62.5 & 1.0\asrdrop{61.5} & 100.0\cacckeep & 98.5\caccdrop{1.5} \\
      & \textit{Phrase}
      & 99.5 & 10.4\asrdrop{89.1} & 92.5\caccdrop{0.4} & 90.4\caccdrop{2.5}
      & 91.8 & 1.2\asrdrop{90.6} & 93.1\caccdrop{2.9} & 92.1\caccdrop{3.9}
      & 91.2 & 0.0\asrdrop{91.2} & 100.0\cacckeep & 100.0\cacckeep \\
      & \textit{Style}
      & 93.8 & 8.8\asrdrop{85.0} & 91.8\caccdrop{1.1} & 89.4\caccdrop{3.5}
      & 96.3 & 2.3\asrdrop{94.0} & 94.4\caccdrop{1.6} & 91.9\caccdrop{4.1}
      & 100.0 & 15.3\asrdrop{84.7} & 99.5\caccdrop{0.5} & 99.8\caccdrop{0.2} \\[0.4ex]
    \midrule

    \rowcolor{gray!20}
    \multicolumn{14}{c}{\bfseries\textit{Qwen2.5-7B-Instruct}} \\
    \cmidrule{1-14}

    \rowcolor{gray!10}
    \textit{Benign}
      & --
      & -- & -- & 92.3 & --
      & -- & -- & 96.8 & --
      & -- & -- & 99.7 & -- \\

    \multirow{3}{*}{\textit{Dirty-label}}
      & \textit{Rare}
      & 100.0 & 6.6\asrdrop{93.4} & 92.4\caccup{0.1} & 89.9\caccdrop{2.4}
      & 100.0 & 8.9\asrdrop{91.1} & 96.3\caccdrop{0.5} & 95.5\caccdrop{1.3}
      & 80.1 & 0.5\asrdrop{79.6} & 100.0\caccup{0.3} & 98.7\caccdrop{1.0} \\
      & \textit{Phrase}
      & 100.0 & 5.8\asrdrop{94.2} & 92.2\caccdrop{0.1} & 90.7\caccdrop{1.6}
      & 100.0 & 4.9\asrdrop{95.1} & 96.4\caccdrop{0.4} & 95.3\caccdrop{1.5}
      & 99.5 & 0.8\asrdrop{98.7} & 100.0\caccup{0.3} & 99.0\caccdrop{0.7} \\
      & \textit{Style}
      & 98.8 & 4.3\asrdrop{94.5} & 92.8\caccup{0.5} & 91.3\caccdrop{1.0}
      & 99.8 & 3.5\asrdrop{96.3} & 96.8\cacckeep & 95.5\caccdrop{1.3}
      & 99.8 & 18.9\asrdrop{80.9} & 99.7\cacckeep & 99.0\caccdrop{0.7} \\[0.4ex]

    \cmidrule(lr){1-2}
    \cmidrule(lr){3-6}
    \cmidrule(lr){7-10}
    \cmidrule(lr){11-14}

    \multirow{3}{*}{\textit{Clean-label}}
      & \textit{Rare}
      & 99.9 & 6.2\asrdrop{93.7} & 91.6\caccdrop{0.7} & 90.8\caccdrop{1.5}
      & 100.0 & 2.8\asrdrop{97.2} & 94.9\caccdrop{1.9} & 95.2\caccdrop{1.6}
      & 71.8 & 2.3\asrdrop{69.5} & 96.6\caccdrop{3.1} & 97.7\caccdrop{2.0} \\
      & \textit{Phrase}
      & 100.0 & 4.6\asrdrop{95.4} & 91.6\caccdrop{0.7} & 89.1\caccdrop{3.2}
      & 96.3 & 3.5\asrdrop{92.8} & 95.4\caccdrop{1.4} & 95.0\caccdrop{1.8}
      & 99.5 & 1.3\asrdrop{98.2} & 100.0\caccup{0.3} & 99.2\caccdrop{0.5} \\
      & \textit{Style}
      & 92.2 & 2.3\asrdrop{89.9} & 91.5\caccdrop{0.8} & 88.9\caccdrop{3.4}
      & 100.0 & 3.7\asrdrop{96.3} & 95.5\caccdrop{1.3} & 95.0\caccdrop{1.8}
      & 99.8 & 13.8\asrdrop{86.0} & 100.0\caccup{0.3} & 99.0\caccdrop{0.7} \\[0.4ex]

    \bottomrule
  \end{tabular}%
  }
\end{table*}

\begin{table*}[t]
  \footnotesize
  \centering
  \caption{\textbf{CC Clamp} results across three models, three datasets, and six attack settings. The \textit{Benign} row corresponds to the clean model without any backdoor attack and therefore reports only clean accuracy (CACC).}
  \label{tab:cc_clamp}

  \setlength{\aboverulesep}{-0.3ex}
  \setlength{\belowrulesep}{0.2ex}
  \renewcommand{\arraystretch}{1.15}

  \resizebox{\textwidth}{!}{%
  \begin{tabular}{cc|cccc|cccc|cccc}
    \toprule

    \multirow{3}{*}{\textbf{Setting}}
      & \multirow{3}{*}{\textbf{Trigger}}
      & \multicolumn{4}{c|}{\textbf{AG News}}
      & \multicolumn{4}{c|}{\textbf{SST-2}}
      & \multicolumn{4}{c}{\textbf{LLM-LAT}} \\

    \cmidrule(lr){3-6}
    \cmidrule(lr){7-10}
    \cmidrule(lr){11-14}

      &
      & \multicolumn{2}{c}{\textbf{ASR}}
      & \multicolumn{2}{c|}{\textbf{CACC}}
      & \multicolumn{2}{c}{\textbf{ASR}}
      & \multicolumn{2}{c|}{\textbf{CACC}}
      & \multicolumn{2}{c}{\textbf{ASR}}
      & \multicolumn{2}{c}{\textbf{CACC}} \\

    \cmidrule(lr){3-4}
    \cmidrule(lr){5-6}
    \cmidrule(lr){7-8}
    \cmidrule(lr){9-10}
    \cmidrule(lr){11-12}
    \cmidrule(lr){13-14}

      &
      & \textit{Before} & \textit{After} & \textit{Before} & \textit{After}
      & \textit{Before} & \textit{After} & \textit{Before} & \textit{After}
      & \textit{Before} & \textit{After} & \textit{Before} & \textit{After} \\
    \midrule

    \rowcolor{gray!20}
    \multicolumn{14}{c}{\bfseries\textit{Gemma-2-9B-IT}} \\
    \cmidrule{1-14}

    \rowcolor{gray!10}
    \textit{Benign}
      & --
      & -- & -- & 93.3 & --
      & -- & -- & 96.6 & --
      & -- & -- & 100.0 & -- \\

    \multirow{3}{*}{\textit{Dirty-label}}
      & \textit{Rare}
      & 100.0 & 0.6\asrdrop{99.4} & 93.0\caccdrop{0.3} & 94.0\caccup{0.7}
      & 100.0 & 0.9\asrdrop{99.1} & 96.0\caccdrop{0.6} & 96.7\caccup{0.1}
      & 73.8 & 9.1\asrdrop{64.7} & 100.0\cacckeep & 100.0\cacckeep \\
      & \textit{Phrase}
      & 100.0 & 0.7\asrdrop{99.3} & 93.5\caccup{0.2} & 93.8\caccup{0.5}
      & 100.0 & 0.9\asrdrop{99.1} & 96.6\cacckeep & 96.6\cacckeep
      & 98.2 & 0.5\asrdrop{97.7} & 100.0\cacckeep & 100.0\cacckeep \\
      & \textit{Style}
      & 98.8 & 3.9\asrdrop{94.9} & 93.3\cacckeep & 93.5\caccup{0.2}
      & 100.0 & 2.6\asrdrop{97.4} & 96.4\caccdrop{0.2} & 96.7\caccup{0.1}
      & 99.5 & 10.8\asrdrop{88.7} & 100.0\cacckeep & 100.0\cacckeep \\[0.4ex]

    \cmidrule(lr){1-2}
    \cmidrule(lr){3-6}
    \cmidrule(lr){7-10}
    \cmidrule(lr){11-14}

    \multirow{3}{*}{\textit{Clean-label}}
      & \textit{Rare}
      & 97.9 & 1.7\asrdrop{96.2} & 92.2\caccdrop{1.1} & 92.6\caccdrop{0.7}
      & 100.0 & 1.0\asrdrop{99.0} & 95.8\caccdrop{0.8} & 96.4\caccdrop{0.2}
      & 82.4 & 1.6\asrdrop{80.8} & 99.8\caccdrop{0.2} & 100.0\cacckeep \\
      & \textit{Phrase}
      & 99.9 & 0.4\asrdrop{99.5} & 92.3\caccdrop{1.0} & 93.0\caccdrop{0.3}
      & 96.7 & 0.9\asrdrop{95.8} & 95.3\caccdrop{1.3} & 96.8\caccup{0.2}
      & 98.2 & 1.5\asrdrop{96.7} & 100.0\cacckeep & 100.0\cacckeep \\
      & \textit{Style}
      & 94.6 & 0.7\asrdrop{93.9} & 92.5\caccdrop{0.8} & 93.7\caccup{0.4}
      & 92.1 & 2.6\asrdrop{89.5} & 96.3\caccdrop{0.3} & 96.7\caccup{0.1}
      & 100.0 & 15.4\asrdrop{84.6} & 99.7\caccdrop{0.3} & 100.0\cacckeep \\[0.4ex]
    \midrule

    \rowcolor{gray!20}
    \multicolumn{14}{c}{\bfseries\textit{Llama-3.1-8B-Instruct}} \\
    \cmidrule{1-14}

    \rowcolor{gray!10}
    \textit{Benign}
      & --
      & -- & -- & 92.9 & --
      & -- & -- & 96.0 & --
      & -- & -- & 100.0 & -- \\

    \multirow{3}{*}{\textit{Dirty-label}}
      & \textit{Rare}
      & 100.0 & 4.5\asrdrop{95.5} & 92.7\caccdrop{0.2} & 92.4\caccdrop{0.5}
      & 100.0 & 4.9\asrdrop{95.1} & 95.9\caccdrop{0.1} & 96.3\caccup{0.3}
      & 61.2 & 0.0\asrdrop{61.2} & 100.0\cacckeep & 99.8\caccdrop{0.2} \\
      & \textit{Phrase}
      & 100.0 & 2.1\asrdrop{97.9} & 93.2\caccup{0.3} & 93.0\caccup{0.1}
      & 100.0 & 3.7\asrdrop{96.3} & 95.9\caccdrop{0.1} & 96.1\caccup{0.1}
      & 64.5 & 0.0\asrdrop{64.5} & 99.7\caccdrop{0.3} & 100.0\cacckeep \\
      & \textit{Style}
      & 99.2 & 5.5\asrdrop{93.7} & 93.2\caccup{0.3} & 92.7\caccdrop{0.2}
      & 99.8 & 3.3\asrdrop{96.5} & 96.0\cacckeep & 96.2\caccup{0.2}
      & 90.2 & 7.8\asrdrop{82.4} & 99.7\caccdrop{0.3} & 100.0\cacckeep \\[0.4ex]

    \cmidrule(lr){1-2}
    \cmidrule(lr){3-6}
    \cmidrule(lr){7-10}
    \cmidrule(lr){11-14}

    \multirow{3}{*}{\textit{Clean-label}}
      & \textit{Rare}
      & 98.3 & 6.2\asrdrop{92.1} & 92.1\caccdrop{0.8} & 93.0\caccup{0.1}
      & 100.0 & 11.9\asrdrop{88.1} & 95.9\caccdrop{0.1} & 96.2\caccup{0.2}
      & 62.5 & 0.0\asrdrop{62.5} & 100.0\cacckeep & 99.8\caccdrop{0.2} \\
      & \textit{Phrase}
      & 99.5 & 3.9\asrdrop{95.6} & 92.5\caccdrop{0.4} & 92.8\caccdrop{0.1}
      & 91.8 & 3.0\asrdrop{88.8} & 93.1\caccdrop{2.9} & 96.4\caccup{0.4}
      & 91.2 & 0.0\asrdrop{91.2} & 100.0\cacckeep & 100.0\cacckeep \\
      & \textit{Style}
      & 93.8 & 1.9\asrdrop{91.9} & 91.8\caccdrop{1.1} & 92.8\caccdrop{0.1}
      & 96.3 & 3.5\asrdrop{92.8} & 94.4\caccdrop{1.6} & 96.1\caccup{0.1}
      & 100.0 & 15.4\asrdrop{84.6} & 99.5\caccdrop{0.5} & 100.0\cacckeep \\[0.4ex]
    \midrule

    \rowcolor{gray!20}
    \multicolumn{14}{c}{\bfseries\textit{Qwen2.5-7B-Instruct}} \\
    \cmidrule{1-14}

    \rowcolor{gray!10}
    \textit{Benign}
      & --
      & -- & -- & 92.3 & --
      & -- & -- & 96.8 & --
      & -- & -- & 99.7 & -- \\

    \multirow{3}{*}{\textit{Dirty-label}}
      & \textit{Rare}
      & 100.0 & 3.1\asrdrop{96.9} & 92.4\caccup{0.1} & 92.3\cacckeep
      & 100.0 & 4.6\asrdrop{95.4} & 96.3\caccdrop{0.5} & 96.9\caccup{0.1}
      & 80.1 & 0.2\asrdrop{79.9} & 100.0\caccup{0.3} & 99.5\caccdrop{0.2} \\
      & \textit{Phrase}
      & 100.0 & 2.2\asrdrop{97.8} & 92.2\caccdrop{0.1} & 92.6\caccup{0.3}
      & 100.0 & 6.1\asrdrop{93.9} & 96.4\caccdrop{0.4} & 96.9\caccup{0.1}
      & 99.5 & 0.2\asrdrop{99.3} & 100.0\caccup{0.3} & 100.0\caccup{0.3} \\
      & \textit{Style}
      & 98.8 & 3.9\asrdrop{94.9} & 92.8\caccup{0.5} & 92.3\cacckeep
      & 99.8 & 5.4\asrdrop{94.4} & 96.8\cacckeep & 96.9\caccup{0.1}
      & 99.8 & 9.1\asrdrop{90.7} & 99.7\cacckeep & 99.8\caccup{0.1} \\[0.4ex]

    \cmidrule(lr){1-2}
    \cmidrule(lr){3-6}
    \cmidrule(lr){7-10}
    \cmidrule(lr){11-14}

    \multirow{3}{*}{\textit{Clean-label}}
      & \textit{Rare}
      & 99.9 & 2.1\asrdrop{97.8} & 91.6\caccdrop{0.7} & 91.5\caccdrop{0.8}
      & 100.0 & 4.7\asrdrop{95.3} & 94.9\caccdrop{1.9} & 96.3\caccdrop{0.5}
      & 71.8 & 0.5\asrdrop{71.3} & 95.6\caccdrop{4.1} & 99.2\caccdrop{0.5} \\
      & \textit{Phrase}
      & 100.0 & 2.7\asrdrop{97.3} & 91.6\caccdrop{0.7} & 92.3\cacckeep
      & 96.3 & 4.4\asrdrop{91.9} & 95.4\caccdrop{1.4} & 96.4\caccdrop{0.4}
      & 99.5 & 0.5\asrdrop{99.0} & 100.0\caccup{0.3} & 99.8\caccup{0.1} \\
      & \textit{Style}
      & 92.2 & 1.8\asrdrop{90.4} & 91.5\caccdrop{0.8} & 92.1\caccdrop{0.2}
      & 100.0 & 5.4\asrdrop{94.6} & 95.5\caccdrop{1.3} & 96.7\caccdrop{0.1}
      & 99.8 & 5.6\asrdrop{94.2} & 100.0\caccup{0.3} & 100.0\caccup{0.3} \\[0.4ex]

    \bottomrule
  \end{tabular}%
  }
\end{table*}

We therefore attribute the interaction-dominated logit shift to SAE features and categorize the top-\(K\) attribution features using \autoref{alg:feature_typing}. As shown in \autoref{tab:four_types}, high-attribution features exhibit distinct four-condition activation profiles. \textsc{Interaction} features remain weak under CC, CT, and PC but sharply activate under PT, indicating features that are specifically recruited only when triggered inputs are processed by the poisoned model. \textsc{Suppressed} features show the opposite pattern, which remain active under normal conditions but collapse under PT, suggesting source-class functionality that is suppressed during backdoor activation. \textsc{Mixed} features are already active under clean conditions and further amplified under PT, reflecting benign and backdoor-related computations that are entangled in the same feature. \textsc{Weight-modified} features are elevated under both PC and PT, capturing persistent representation changes introduced by poisoned training.

\subsubsection{Feature-Level Encoding Divergence Drives Defense Fragmentation}
\label{sec:divergence}
The feature types defined above describe attack-conditioned roles rather than fixed semantic labels. 
We therefore compare their composition across attack paradigms to examine whether dirty-label and clean-label backdoors rely on different feature-level encoding strategies.

As shown in \autoref{tab:divergence}, dirty-label and clean-label attacks exhibit clearly different compositions. 
Dirty-label attacks are consistently dominated by \textsc{Interaction} features, which account for at least \textbf{27} out of 60 top-attributed features across the three trigger types. These features remain inactive or weakly activated under the trigger-only and weight-only conditions, but become strongly activated when the trigger and poisoned weights are jointly present. This indicates that dirty-label backdoors are primarily encoded through isolated trigger--weight interaction mechanisms.

Clean-label attacks show a different pattern. Compared with dirty-label attacks, they contain substantially fewer \textsc{Interaction} features and are instead dominated by \textsc{Mixed} features, which account for at least \textbf{23} out of 60 top-attributed features across all clean-label settings. They also contain more \textsc{Weight-modified} features than dirty-label attacks. This suggests that clean-label backdoors are less likely to be implemented as isolated backdoor components; rather, their backdoor-relevant signals are more often embedded into features that also participate in normal model computation or reflect persistent weight-level shifts introduced during poisoned training.

This divergence explains why existing defenses struggle to provide unified protection across both paradigms. Defenses built on the assumption that backdoor behavior is concentrated in separable trigger-induced components, such as BEEAR and CRoW, align well with the interaction-dominated structure of dirty-label attacks. However, this assumption becomes less reliable for clean-label attacks, where backdoor-relevant features are more frequently entangled with benign functionality. Conversely, methods that aim to regularize, decouple, or reshape internal representations, such as RepGuard and DeCE, are better aligned with the entangled nature of clean-label attacks, but they do not directly target the sharp interaction features that dominate dirty-label backdoors.

The fragmentation is therefore structural rather than methodological: \textit{dirty-label and clean-label attacks exploit different feature-level encoding strategies, making single fixed defensive assumption unlikely to generalize across both paradigms.}

\section{Mechanism-Guided Validation of Backdoor Mitigation}
\label{sec:clamp}
The analyses in \S\ref{sec:mechanisms} suggest that backdoor behaviors are mediated by identifiable SAE features whose compositions differ across attack paradigms. We validate this claim through causal intervention: if these features capture the underlying mechanism, then realigning them should mitigate the attack. To keep the validation minimal, we simply clamp suspicious SAE feature activations to reference values at inference time.

\subsection{Experimental Setup}
We use the same setup as in \S\ref{sec:motivation}, covering three architectures, three datasets, and six attack settings. For each backdoored model, we select the top-\(k\) suspicious SAE features by ranking their statistical divergence between clean and suspicious inputs. Full details are provided in \autoref{app:experiment setup}.

We evaluate two reference sources: \textbf{Base Clamp}, which uses the pretrained base model, and \textbf{CC Clamp}, which uses a clean fine-tuned model. For each method, we report the lowest ASR achieved under minimal CACC degradation.

\subsection{Suspicious Feature Selection}
\label{sec:suspicious_feature_selection}
Notably, clamping uses a different feature selection strategy from the attribution-based analysis in \S~\ref{sec:framework}, which assumes known backdoor behavior and relies on logit attribution and $2\times2$ activation patterns. Here, we target a more realistic setting where the defender knows neither the trigger nor the target. We therefore rank SAE features by how strongly their activations on protected inputs deviate from a small clean calibration set, and select the top-$k$ features for intervention. Details are provided in \autoref{app:suspicious_feature_selection}.

\subsection{Main Results}
\autoref{tab:base_clamp} and \autoref{tab:cc_clamp} report the Base Clamp and CC Clamp results. Both variants substantially reduce ASR in most settings while preserving clean accuracy, showing that feature-level realignment effectively mitigates backdoor behaviors.

\subsubsection{Clamp Effectiveness.}
\paragraph{Base Clamp.}
Feature-level clamping consistently suppresses backdoor behavior across models, datasets, and attack settings. Even using only the pretrained base model as reference, \textbf{Base Clamp} substantially reduces ASR in nearly all configurations. On AG News and SST-2, post-clamping ASR is generally reduced to \textbf{below 10\%}, with only a few clean-label cases on Llama-3.1-8B-Instruct reaching at most 15\%. On LLM-LAT, Base Clamp also achieves large ASR reductions, especially against phrase-trigger attacks, although style-trigger attacks and several clean-label cases remain more challenging, with residual ASR reaching at most 28\%. CACC is largely preserved across nearly all configurations.

\paragraph{CC Clamp.}
Using a clean fine-tuned model as reference further strengthens the intervention. Across all configurations, \textbf{CC Clamp} reduces ASR to at most \textbf{15.4\%}, and in most cases to \textbf{below 6\%}. The improvement is particularly pronounced on LLM-LAT: across all three architectures, phrase-trigger attacks are almost completely neutralized, and rare-token attacks are reduced to \textbf{near-zero} ASR in most settings. CACC remains largely unchanged across nearly all configurations, often matching or slightly exceeding the original clean accuracy after intervention.

\begin{figure}[htb]
  \centering
  \includegraphics[width=\columnwidth]{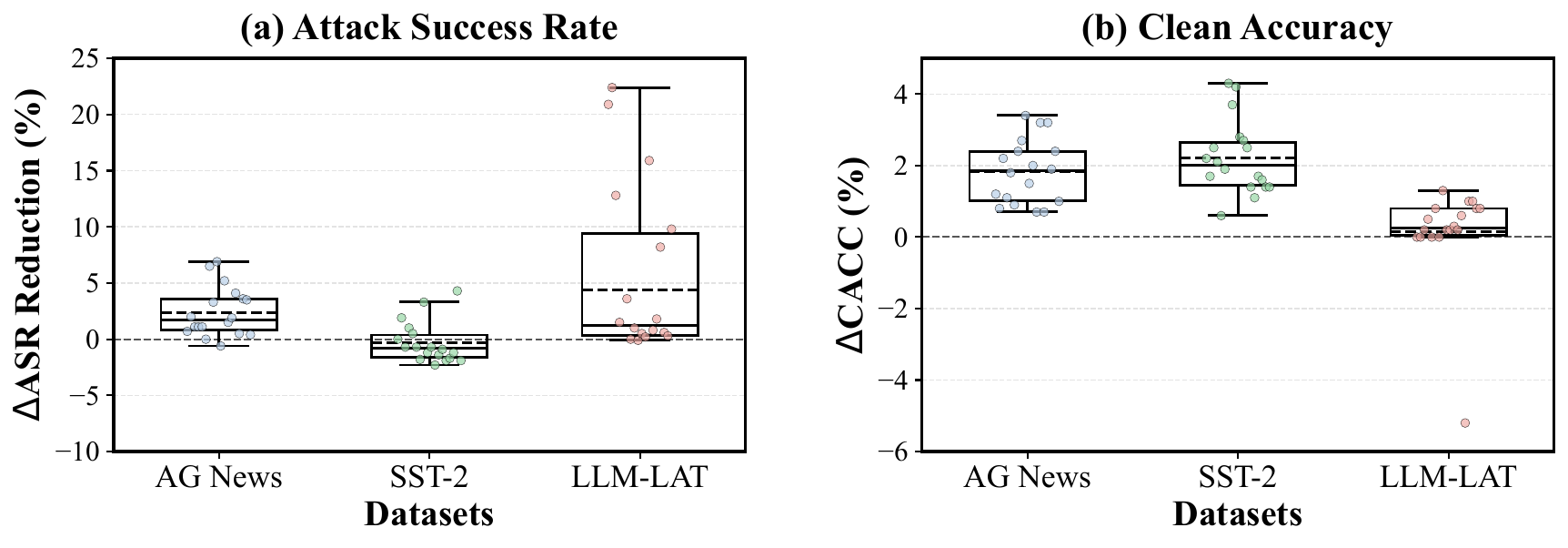}
  \caption{Results of $\Delta ASR$ and $\Delta CACC$ between Base Clamp and CC Clamp.}
  \label{fig:main_asr_deltaCACC}
\end{figure}

\paragraph{Base Clamp vs. CC Clamp.}
CC Clamp achieves stronger mitigation than Base Clamp, but the gap is modest in most settings. As shown in \autoref{fig:main_asr_deltaCACC}, ASR differences are concentrated near zero on AG News and SST-2, indicating that the pretrained base model already provides an effective reference for feature realignment. LLM-LAT shows larger variance, where a clean fine-tuned reference brings additional gains in some safety-alignment cases. For CACC, the differences are smaller, with most values close to zero across all datasets. Thus, while CC Clamp is preferable when a clean fine-tuned model is available, Base Clamp remains a practical alternative that requires only the publicly available pretrained base model.

\subsubsection{Comparison with Existing Defenses}
\begin{figure}[htb]
  \centering
  \includegraphics[width=\columnwidth]{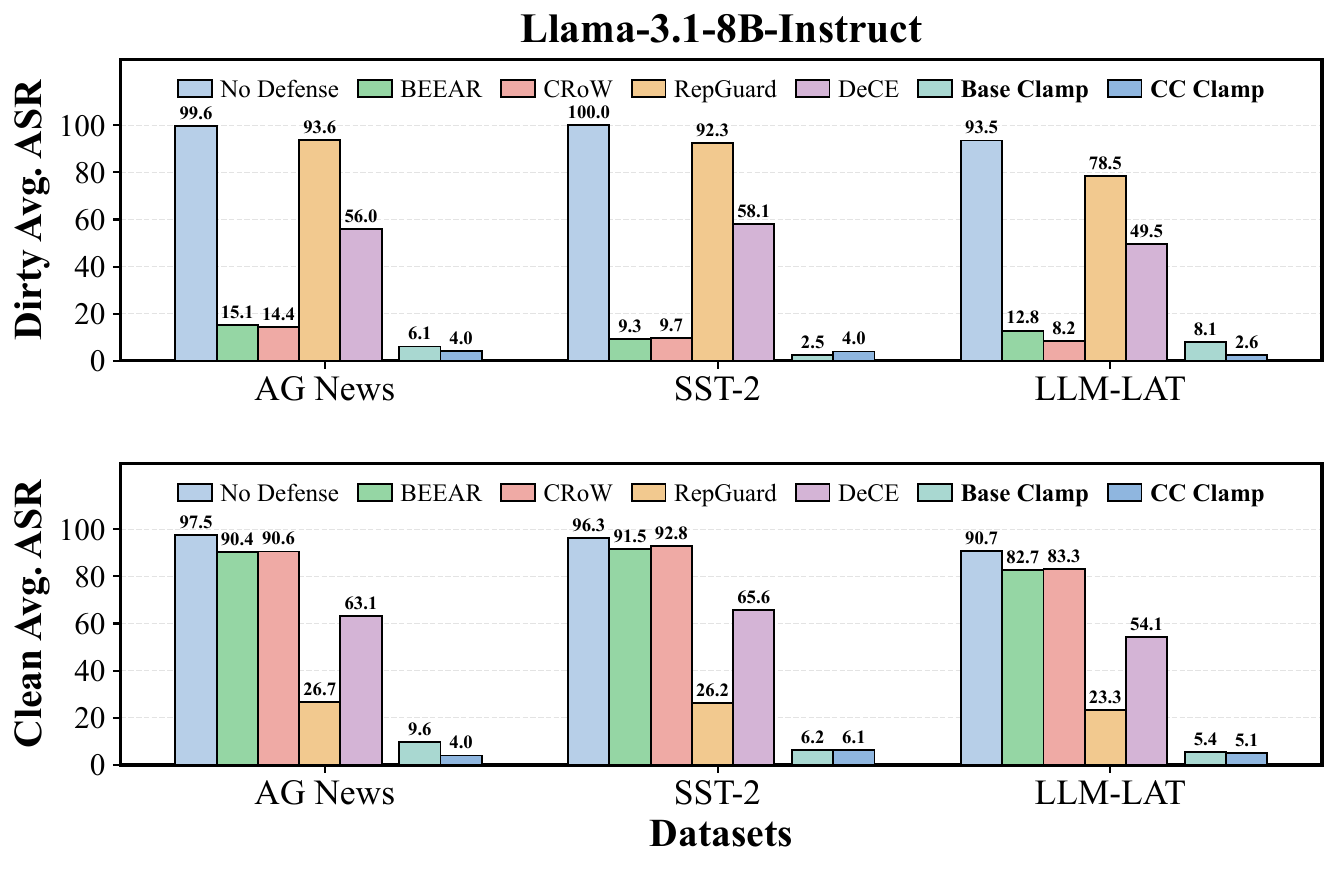}
  \caption{Results of Avg. ASR comparison between baselines and clamping on Llama-3.1-8B-Instruct.}
  \label{fig:baseline_comparison}
\end{figure}

To contextualize the effectiveness of feature-level clamping, we compare Base Clamp and CC Clamp with four representative defenses on Llama-3.1-8B-Instruct, with results on the other two model architectures provided in \autoref{app:full_result_of_comparison_with_existing_defense}. As shown in \autoref{fig:baseline_comparison}, existing defenses exhibit clear paradigm-specific limitations: BEEAR and CRoW are effective under dirty-label attacks but fail to generalize to clean-label settings, whereas RepGuard shows the opposite trend, substantially reducing clean-label ASR while remaining largely ineffective against dirty-label attacks. DeCE provides moderate mitigation in both settings but still leaves relatively high residual ASR. By contrast, both Base Clamp and CC Clamp achieve consistently low ASR across the two attack paradigms. Base Clamp reduces the average ASR to at most 8.1\% under dirty-label attacks and 9.6\% under clean-label attacks, while CC Clamp further reduces it to at most 6.1\% and 6.2\%, respectively. This consistent performance suggests that feature-level clamping directly targets the SAE features mediating backdoor behavior, enabling unified mitigation across heterogeneous backdoor mechanisms.

\section{Conclusion}
In this work, we present the first systematic feature-level mechanistic analysis of LLM backdoors using sparse autoencoders. By tracing backdoor-induced logit shifts to high-contributing SAE features, we identify four functional feature roles and reveal systematic encoding differences between dirty-label and clean-label attacks. These differences provide a mechanistic explanation for why existing defenses remain fragmented across attack paradigms. Inference-time feature clamping further validates the causal role of these features and demonstrates their potential for interpretable backdoor mitigation while largely preserving benign performance.

\clearpage
\section*{Limitations}

\paragraph{Datasets and Settings.}
Although our experiments cover multiple model families, datasets, trigger types, and both dirty-label and clean-label attacks, they still represent a finite subset of possible backdoor scenarios. In particular, our evaluation focuses on controlled classification-style settings and a limited set of backdoor objectives. Future work could examine whether the observed feature-level mechanisms generalize to larger-scale models, open-ended generation tasks, multi-trigger attacks, and more adaptive poisoning strategies.

\paragraph{Potential Risks.}
By exposing feature-level mechanisms underlying backdoor activation, our analysis could also provide insights that help adaptive attackers design more evasive backdoors. For example, attackers may try to distribute backdoor behavior across less salient features or make it more entangled with benign functionality. We therefore view SAE-based analysis as a tool that should be used carefully, with emphasis on diagnosis, auditing, and mitigation rather than attack optimization.

\paragraph{Defense Scope.}
Our inference-time feature clamping experiments are primarily designed to validate the causal role of identified SAE features, rather than to serve as a fully optimized deployment-time defense. The current procedure still requires feature selection and calibration, and its efficiency may depend on the SAE size and intervention layer. Building more automated, lightweight, and robust SAE-based defenses is an important direction for future work.
\section*{Acknowledgments}
This work was supported by the Beijing High Innovation Plan (No. 202504841069). The authors gratefully acknowledge the financial support provided by this program. We also wish to express our sincere gratitude to all the anonymous reviewers for their valuable comments and constructive suggestions, which have significantly improved the quality of this paper.

\vspace{-1em}

\bibliography{custom}

@article{yu2025backdoor,
  title={Backdoor Attribution: Elucidating and Controlling Backdoor in Language Models},
  author={Yu, Miao and Zhou, Zhenhong and Aloqaily, Moayad and Wang, Kun and Huang, Biwei and Wang, Stephen and Jin, Yueming and Wen, Qingsong},
  journal={arXiv preprint arXiv:2509.21761},
  year={2025}
}

@inproceedings{yi2024badacts,
  title={BadActs: A universal backdoor defense in the activation space},
  author={Yi, Biao and Chen, Sishuo and Li, Yiming and Li, Tong and Zhang, Baolei and Liu, Zheli},
  booktitle={Findings of the Association for Computational Linguistics: ACL 2024},
  pages={5339--5352},
  year={2024}
}

@inproceedings{zhao2024defense,
  title={Defense against backdoor attack on pre-trained language models via head pruning and attention normalization},
  author={Zhao, Xingyi and Xu, Depeng and Yuan, Shuhan},
  booktitle={Forty-first International Conference on Machine Learning},
  year={2024}
}

@article{niu2026repguard,
  title={RepGuard: Adaptive Feature Decoupling for Robust Backdoor Defense in Large Language Models},
  author={Niu, Chenxu and Zhang, Jie and Liu, Yanbing and Li, Yunpeng and Weng, Jinta and Hu, Yue},
  journal={Advances in Neural Information Processing Systems},
  volume={38},
  pages={44851--44872},
  year={2026}
}

@article{assogba2026sparse,
  title={Sparse Autoencoders are Capable LLM Jailbreak Mitigators},
  author={Assogba, Yannick and Cortellazzi, Jacopo and Abad, Javier and Rodriguez, Pau and Suau, Xavier and Blaas, Arno},
  journal={arXiv preprint arXiv:2602.12418},
  year={2026}
}

@article{cunningham2023sparse,
  title={Sparse autoencoders find highly interpretable features in language models},
  author={Cunningham, Hoagy and Ewart, Aidan and Riggs, Logan and Huben, Robert and Sharkey, Lee},
  journal={arXiv preprint arXiv:2309.08600},
  year={2023}
}

@inproceedings{gan2022triggerless,
  title={Triggerless backdoor attack for NLP tasks with clean labels},
  author={Gan, Leilei and Li, Jiwei and Zhang, Tianwei and Li, Xiaoya and Meng, Yuxian and Wu, Fei and Yang, Yi and Guo, Shangwei and Fan, Chun},
  booktitle={Proceedings of the 2022 Conference of the North American Chapter of the Association for Computational Linguistics: Human Language Technologies},
  pages={2942--2952},
  year={2022}
}

@article{gu2017badnets,
  title={Badnets: Identifying vulnerabilities in the machine learning model supply chain},
  author={Gu, Tianyu and Dolan-Gavitt, Brendan and Garg, Siddharth},
  journal={arXiv preprint arXiv:1708.06733},
  year={2017}
}

@inproceedings{chen2021badnl,
  title={Badnl: Backdoor attacks against nlp models with semantic-preserving improvements},
  author={Chen, Xiaoyi and Salem, Ahmed and Chen, Dingfan and Backes, Michael and Ma, Shiqing and Shen, Qingni and Wu, Zhonghai and Zhang, Yang},
  booktitle={Proceedings of the 37th Annual Computer Security Applications Conference},
  pages={554--569},
  year={2021}
}

@inproceedings{zeng2024beear,
  title={Beear: Embedding-based adversarial removal of safety backdoors in instruction-tuned language models},
  author={Zeng, Yi and Sun, Weiyu and Huynh, Tran and Song, Dawn and Li, Bo and Jia, Ruoxi},
  booktitle={Proceedings of the 2024 Conference on Empirical Methods in Natural Language Processing},
  pages={13189--13215},
  year={2024}
}

@inproceedings{qi2021hidden,
  title={Hidden killer: Invisible textual backdoor attacks with syntactic trigger},
  author={Qi, Fanchao and Li, Mukai and Chen, Yangyi and Zhang, Zhengyan and Liu, Zhiyuan and Wang, Yasheng and Sun, Maosong},
  booktitle={Proceedings of the 59th Annual Meeting of the Association for Computational Linguistics and the 11th International Joint Conference on Natural Language Processing (Volume 1: Long Papers)},
  pages={443--453},
  year={2021}
}

@inproceedings{min2025crow,
    title={{CROW}: Eliminating Backdoors from Large Language Models via Internal Consistency Regularization},
    author={Nay Myat Min and Long H. Pham and Yige Li and Jun Sun},
    booktitle={Forty-second International Conference on Machine Learning},
    year={2025}
}

@article{yang2026defending,
  title={Defending code language models against backdoor attacks with deceptive cross-entropy loss},
  author={Yang, Guang and Zhou, Yu and Zhang, Xiangyu and Chen, Xiang and Zhuo, Terry Yue and Lo, David and Chen, Taolue},
  journal={ACM Transactions on Software Engineering and Methodology},
  volume={35},
  number={2},
  pages={1--27},
  year={2026},
  publisher={ACM New York, NY}
}

@article{team2024gemma,
  title={Gemma 2: Improving open language models at a practical size},
  author={Team, Gemma and Riviere, Morgane and Pathak, Shreya and Sessa, Pier Giuseppe and Hardin, Cassidy and Bhupatiraju, Surya and Hussenot, L{\'e}onard and Mesnard, Thomas and Shahriari, Bobak and Ram{\'e}, Alexandre and others},
  journal={arXiv preprint arXiv:2408.00118},
  year={2024}
}

@article{yang2024qwen25,
  title={Qwen2.5 Technical Report},
  author={Yang, An and Yang, Baosong and Zhang, Beichen and Hui, Binyuan and Zheng, Bo and Yu, Bowen and Li, Chengyuan and Liu, Dayiheng and Huang, Fei and Wei, Haoran and others},
  journal={arXiv preprint arXiv:2412.15115},
  year={2024}
}

@article{grattafiori2024llama,
  title={The llama 3 herd of models},
  author={Grattafiori, Aaron and Dubey, Abhimanyu and Jauhri, Abhinav and Pandey, Abhinav and Kadian, Abhishek and Al-Dahle, Ahmad and Letman, Aiesha and Mathur, Akhil and Schelten, Alan and Vaughan, Alex and others},
  journal={arXiv preprint arXiv:2407.21783},
  year={2024}
}

@article{zhang2015character,
  title={Character-level convolutional networks for text classification},
  author={Zhang, Xiang and Zhao, Junbo and LeCun, Yann},
  journal={Advances in neural information processing systems},
  volume={28},
  year={2015}
}

@article{sheshadri2024targeted,
  title={Targeted Latent Adversarial Training Improves Robustness to Persistent Harmful Behaviors in LLMs},
  author={Sheshadri, Abhay and Ewart, Aidan and Guo, Phillip and Lynch, Aengus and Wu, Cindy and Hebbar, Vivek and Sleight, Henry and Stickland, Asa Cooper and Perez, Ethan and Hadfield-Menell, Dylan and Casper, Stephen},
  journal={arXiv preprint arXiv:2407.15549},
  year={2024}
}

@inproceedings{socher2013recursive,
  title={Recursive deep models for semantic compositionality over a sentiment treebank},
  author={Socher, Richard and Perelygin, Alex and Wu, Jean and Chuang, Jason and Manning, Christopher D and Ng, Andrew Y and Potts, Christopher},
  booktitle={Proceedings of the 2013 conference on empirical methods in natural language processing},
  pages={1631--1642},
  year={2013}
}

@article{hubinger2024sleeper,
  title={Sleeper agents: Training deceptive llms that persist through safety training},
  author={Hubinger, Evan and Denison, Carson and Mu, Jesse and Lambert, Mike and Tong, Meg and MacDiarmid, Monte and Lanham, Tamera and Ziegler, Daniel M and Maxwell, Tim and Cheng, Newton and others},
  journal={arXiv preprint arXiv:2401.05566},
  year={2024}
}

@article{elhage2022superposition,
  title     = {Toy Models of Superposition},
  author    = {Nelson Elhage and Tristan Hume and Catherine Olsson and Nicholas Schiefer and Tom Henighan and Shauna Kravec and Zac Hatfield-Dodds and Robert Lasenby and Dawn Drain and Carol Chen and Roger Grosse and Sam McCandlish and Jared Kaplan and Dario Amodei and Martin Wattenberg and Christopher Olah},
  journal   = {arXiv preprint arXiv:2209.10652},
  year      = {2022}
}

@inproceedings{yan2024backdooring,
  title={Backdooring instruction-tuned large language models with virtual prompt injection},
  author={Yan, Jun and Yadav, Vikas and Li, Shiyang and Chen, Lichang and Tang, Zheng and Wang, Hai and Srinivasan, Vijay and Ren, Xiang and Jin, Hongxia},
  booktitle={Proceedings of the 2024 Conference of the North American Chapter of the Association for Computational Linguistics: Human Language Technologies (Volume 1: Long Papers)},
  pages={6065--6086},
  year={2024}
}

@inproceedings{
tong2025badjudge,
title={BadJudge: Backdoor Vulnerabilities of {LLM}-As-A-Judge},
author={Terry Tong and Fei Wang and Zhe Zhao and Muhao Chen},
booktitle={The Thirteenth International Conference on Learning Representations},
year={2025}
}

@inproceedings{kurita2020weight,
  title={Weight poisoning attacks on pretrained models},
  author={Kurita, Keita and Michel, Paul and Neubig, Graham},
  booktitle={Proceedings of the 58th annual meeting of the association for computational linguistics},
  pages={2793--2806},
  year={2020}
}

@inproceedings{gao2019strip,
  title={Strip: A defence against trojan attacks on deep neural networks},
  author={Gao, Yansong and Xu, Change and Wang, Derui and Chen, Shiping and Ranasinghe, Damith C and Nepal, Surya},
  booktitle={Proceedings of the 35th annual computer security applications conference},
  pages={113--125},
  year={2019}
}

@article{chen2018detecting,
  title={Detecting backdoor attacks on deep neural networks by activation clustering},
  author={Chen, Bryant and Carvalho, Wilka and Baracaldo, Nathalie and Ludwig, Heiko and Edwards, Benjamin and Lee, Taesung and Molloy, Ian and Srivastava, Biplav},
  journal={arXiv preprint arXiv:1811.03728},
  year={2018}
}

@inproceedings{qi2021onion,
  title={Onion: A simple and effective defense against textual backdoor attacks},
  author={Qi, Fanchao and Chen, Yangyi and Li, Mukai and Yao, Yuan and Liu, Zhiyuan and Sun, Maosong},
  booktitle={Proceedings of the 2021 conference on empirical methods in natural language processing},
  pages={9558--9566},
  year={2021}
}

@inproceedings{yang2021rap,
  title={Rap: Robustness-aware perturbations for defending against backdoor attacks on nlp models},
  author={Yang, Wenkai and Lin, Yankai and Li, Peng and Zhou, Jie and Sun, Xu},
  booktitle={Proceedings of the 2021 Conference on Empirical Methods in Natural Language Processing},
  pages={8365--8381},
  year={2021}
}

@inproceedings{
arditi2024refusal,
title={Refusal in Language Models Is Mediated by a Single Direction},
author={Andy Arditi and Oscar Balcells Obeso and Aaquib Syed and Daniel Paleka and Nina Rimsky and Wes Gurnee and Neel Nanda},
booktitle={The Thirty-eighth Annual Conference on Neural Information Processing Systems},
year={2024}
}

@inproceedings{
chen2026towards,
title={Towards Understanding Safety Alignment: A Mechanistic Perspective from Safety Neurons},
author={Jianhui Chen and Xiaozhi Wang and Zijun Yao and Yushi Bai and Lei Hou and Juanzi Li},
booktitle={The Thirty-ninth Annual Conference on Neural Information Processing Systems},
year={2026}
}

@article{he2024jailbreaklens,
  title={Jailbreaklens: Interpreting jailbreak mechanism in the lens of representation and circuit},
  author={He, Zeqing and Wang, Zhibo and Chu, Zhixuan and Xu, Huiyu and Zhang, Wenhui and Wang, Qinglong and Zheng, Rui},
  journal={arXiv preprint arXiv:2411.11114},
  year={2024}
}

@article{li2026backdoorllm,
  title={Backdoorllm: A comprehensive benchmark for backdoor attacks and defenses on large language models},
  author={Li, Yige and Huang, Hanxun and Zhao, Yunhan and Ma, Xingjun and Sun, Jun},
  journal={Advances in neural information processing systems},
  volume={38},
  year={2026}
}

@inproceedings{du2024uor,
  title={Uor: Universal backdoor attacks on pre-trained language models},
  author={Du, Wei and Li, Peixuan and Zhao, Haodong and Ju, Tianjie and Ren, Ge and Liu, Gongshen},
  booktitle={Findings of the Association for Computational Linguistics: ACL 2024},
  pages={7865--7877},
  year={2024}
}

@inproceedings{zhou2025role,
  title={On the role of attention heads in large language model safety},
  author={Zhou, Zhenhong and Yu, Haiyang and Zhang, Xinghua and Xu, Rongwu and Huang, Fei and Wang, Kun and Liu, Yang and Fang, Junfeng and Li, Yongbin},
  booktitle={International Conference on Learning Representations},
  volume={2025},
  pages={84042--84071},
  year={2025}
}

@article{wang2025interpretable,
  title={Interpretable Safety Alignment via SAE-Constructed Low-Rank Subspace Adaptation},
  author={Wang, Dianyun and Ma, Qingsen and Shang, Yuhu and Lu, Zhifeng and Xu, Zhenbo and Ning, Lechen and Wu, Huijia and He, Zhaofeng},
  journal={arXiv preprint arXiv:2512.23260},
  year={2025}
}

@article{wang2025enhancing,
  title={Enhancing LLM Steering through Sparse Autoencoder-Based Vector Refinement},
  author={Wang, Anyi and Wu, Xuansheng and Shu, Dong and Ma, Yunpu and Liu, Ninghao},
  journal={arXiv preprint arXiv:2509.23799},
  year={2025}
}

@misc{
cho2026corrsteer,
title={CorrSteer: Generation-Time {LLM} Steering via Correlated Sparse Autoencoder Features},
author={Seonglae Cho and Zekun Wu and Adriano Koshiyama},
year={2026}
}

@article{chen2022effective,
  title={Effective backdoor defense by exploiting sensitivity of poisoned samples},
  author={Chen, Weixin and Wu, Baoyuan and Wang, Haoqian},
  journal={Advances in Neural Information Processing Systems},
  volume={35},
  pages={9727--9737},
  year={2022}
}

@article{Yin_Wang_Lin_Liu_2025, title={Adversarial-Inspired Backdoor Defense via Bridging Backdoor and Adversarial Attacks}, volume={39}, DOI={10.1609/aaai.v39i9.33030}, number={9}, journal={Proceedings of the AAAI Conference on Artificial Intelligence}, author={Yin, Jia-Li and Wang, Weijian and , Lyhwa and Lin, Wei and Liu, Ximeng}, year={2025}, month={Apr.}, pages={9508–9516} }

@inproceedings{zeng2026miragebackdoor,
  title={MirageBackdoor: A Stealthy Attack that Induces Think-Well-Answer-Wrong Reasoning},
  author={Zeng, Yizhe and Zhang, Wei and Li, Yunpeng and Xiao, Juxin and Wang, Xiao and Liu, Yuling},
  booktitle={Proceedings of the 64th Annual Meeting of the Association for Computational Linguistics (Volume 1: Long Papers)},
  pages={8639--8659},
  year={2026}
}

\clearpage
\appendix

\section{Details of ASR and CACC Metrics}
\label{app:metrics}
We evaluate backdoor effectiveness and benign utility using two standard metrics: Attack Success Rate (ASR) and Clean Accuracy (CACC). Let \(F\) denote a clean model and \(F'\) denote the corresponding backdoored model. For a clean input \(x\), let \(\tau(x)\) denote the triggered input obtained by inserting a predefined trigger into \(x\). Let \(y^{\mathrm{atk}}(x)\) be the attacker-specified target output for \(x\).

\paragraph{Attack Success Rate.}
ASR measures how often the backdoored model produces the attacker-specified output on triggered inputs. Given a triggered evaluation distribution \(\mathcal{D}_{\mathrm{trig}}\), ASR is defined as
\begin{equation}
    \mathrm{ASR}(F')
    =
    \mathbb{E}_{x \sim \mathcal{D}_{\mathrm{trig}}}
    \left[
    \mathbf{1}
    \left(
    F'(\tau(x)) = y^{\mathrm{atk}}(x)
    \right)
    \right],
\end{equation}
where \(\mathbf{1}(\cdot)\) is the indicator function. For a finite triggered test set \(\mathcal{T}_{\mathrm{test}}\), we compute the empirical ASR as
\begin{equation}
    \widehat{\mathrm{ASR}}(F')
    =
    \frac{1}{|\mathcal{T}_{\mathrm{test}}|}
    \sum_{x \in \mathcal{T}_{\mathrm{test}}}
    \mathbf{1}
    \left(
    F'(\tau(x)) = y^{\mathrm{atk}}(x)
    \right).
\end{equation}

\paragraph{Clean Accuracy.}
CACC measures whether the backdoored model preserves its normal task performance on clean inputs. Let \(\mathcal{D}_{\mathrm{clean}}\) denote the clean test distribution, where each example is paired with its ground-truth label or answer \(y\). CACC is defined as
\begin{equation}
    \mathrm{CACC}(F')
    =
    \mathbb{E}_{(x,y) \sim \mathcal{D}_{\mathrm{clean}}}
    \left[
    \mathbf{1}
    \left(
    F'(x)=y
    \right)
    \right].
\end{equation}
Given a finite clean test set \(\mathcal{D}_{\mathrm{test}}\), the empirical estimate is
\begin{equation}
    \widehat{\mathrm{CACC}}(F')
    =
    \frac{1}{|\mathcal{D}_{\mathrm{test}}|}
    \sum_{(x,y) \in \mathcal{D}_{\mathrm{test}}}
    \mathbf{1}
    \left(
    F'(x)=y
    \right).
\end{equation}
Intuitively, ASR reflects the strength of the implanted backdoor under trigger activation, while CACC reflects the stealthiness of the attack by measuring whether the model retains normal performance on benign inputs.

\section{Full Results of Defense Fragmentation}
\label{app:motivation_full_results}
In this section, we report the detailed cross-evaluation results on Qwen2.5-7B-Instruct and Gemma-2-9B-IT. These results complement the Llama-3.1-8B-Instruct results in \autoref{tab:motivation} and provide a broader view of defense fragmentation across different model families.

\paragraph{Results on Qwen2.5-7B-Instruct.}
\autoref{tab:motivation_qwen} reports the full cross-evaluation results on Qwen2.5-7B-Instruct. The results show the same asymmetric pattern as observed on Llama-3.1-8B-Instruct. BEEAR and CRoW effectively mitigate dirty-label backdoors, reducing the average ASR to 10.6\% and 10.1\%, respectively, across datasets and triggers. However, they remain much less effective under clean-label attacks, where the average ASR stays high at 84.7\% and 85.2\%. RepGuard exhibits the opposite behavior: it reduces clean-label ASR to 22.4\% on average, but leaves dirty-label attacks largely unaffected, with an average ASR of 84.8\%. DeCE provides more balanced mitigation across the two paradigms, but its remaining ASR is still substantially higher than the best paradigm-specific defenses, averaging 51.8\% under dirty-label attacks and 57.5\% under clean-label attacks. These results indicate that even on Qwen2.5-7B-Instruct, where defenses are generally more effective, no method achieves consistent protection across both dirty-label and clean-label backdoors.

\paragraph{Results on Gemma-2-9B-IT.}
\autoref{tab:motivation_gemma} reports the corresponding results on Gemma-2-9B-IT. The same fragmentation pattern persists. BEEAR and CRoW strongly suppress dirty-label attacks, reducing the average ASR to 13.9\% and 13.3\%, respectively, but they fail to generalize to clean-label attacks, where the average ASR remains 87.0\% and 87.6\%. In contrast, RepGuard substantially reduces clean-label ASR to 26.8\% on average, while dirty-label ASR remains high at 87.4\%. DeCE again provides moderate but incomplete mitigation under both settings, with average ASR of 56.4\% for dirty-label attacks and 60.6\% for clean-label attacks. 

Together with the Llama results, these results confirm that defense fragmentation is not specific to a single model family. Existing defenses remain strongly tied to paradigm-specific assumptions, showing strong performance only when the attack mechanism matches the defense design.
\begin{table}[t]
    \centering
    \small
    \caption{Cross-evaluation results of existing defenses on Qwen2.5-7B-Instruct.}
    \label{tab:motivation_qwen}
    \setlength{\tabcolsep}{6pt}
    \renewcommand{\arraystretch}{1.15}
    
    \resizebox{\columnwidth}{!}{%
    \begin{tabular}{cc ccc ccc}
        \toprule
        
        \multirow{2}{*}{\textbf{Dataset}} 
        & \multirow{2}{*}{\textbf{Defense}}
        & \multicolumn{3}{c}{\textbf{Dirty-label ASR (\%)}} 
        & \multicolumn{3}{c}{\textbf{Clean-label ASR (\%)}} \\
        
        \cmidrule(lr){3-5}
        \cmidrule(lr){6-8}
        
        & 
        & \textbf{Rare}
        & \textbf{Phrase}
        & \textbf{Style}
        & \textbf{Rare}
        & \textbf{Phrase}
        & \textbf{Style} \\
        
        \midrule
        
        \multirow{5}{*}{AG News}
        & No Defense & 100 & 100 & 98.8 & 97.9 & 99.9 & 94.6 \\
        & BEEAR      & 7.4 & 10.2 & 15.8 & 86.5 & 88.7 & 90.9 \\
        & CRoW       & 8.9 & 11.4 & 14.6 & 85.8 & 88.1 & 91.0 \\
        & RepGuard   & 88.6 & 91.0 & 94.5 & 18.7 & 23.4 & 29.8 \\
        & DeCE       & 48.6 & 52.7 & 57.9 & 54.1 & 58.6 & 63.7 \\
        
        \midrule
        
        \multirow{5}{*}{SST-2}
        & No Defense & 100 & 100 & 100 & 100 & 96.7 & 92.1 \\
        & BEEAR      & 4.9 & 7.8 & 12.6 & 88.9 & 91.0 & 89.8 \\
        & CRoW       & 5.8 & 8.2 & 13.1 & 89.4 & 90.8 & 92.5 \\
        & RepGuard   & 89.8 & 91.7 & 93.4 & 15.9 & 21.8 & 27.6 \\
        & DeCE       & 50.7 & 54.3 & 58.6 & 57.5 & 62.1 & 65.8 \\
        
        \midrule
        
        \multirow{5}{*}{LLM-LAT}
        & No Defense & 82.4 & 98.2 & 100 & 74.3 & 98.2 & 99.5 \\
        & BEEAR      & 8.6 & 11.9 & 16.5 & 61.7 & 85.6 & 89.2 \\
        & CRoW       & 6.4 & 9.5 & 12.8 & 63.1 & 86.4 & 90.0 \\
        & RepGuard   & 73.8 & 77.9 & 82.6 & 15.8 & 21.7 & 26.9 \\
        & DeCE       & 41.5 & 47.9 & 53.6 & 44.8 & 52.6 & 58.7 \\
        
        \bottomrule
    \end{tabular}%
    }
\end{table}

\begin{table}[t]
    \centering
    \small
    \caption{Cross-evaluation results of existing defenses on Gemma-2-9B-IT.}
    \label{tab:motivation_gemma}
    \setlength{\tabcolsep}{6pt}
    \renewcommand{\arraystretch}{1.15}
    
    \resizebox{\columnwidth}{!}{%
    \begin{tabular}{cc ccc ccc}
        \toprule
        
        \multirow{2}{*}{\textbf{Dataset}} 
        & \multirow{2}{*}{\textbf{Defense}}
        & \multicolumn{3}{c}{\textbf{Dirty-label ASR (\%)}} 
        & \multicolumn{3}{c}{\textbf{Clean-label ASR (\%)}} \\
        
        \cmidrule(lr){3-5}
        \cmidrule(lr){6-8}
        
        & 
        & \textbf{Rare}
        & \textbf{Phrase}
        & \textbf{Style}
        & \textbf{Rare}
        & \textbf{Phrase}
        & \textbf{Style} \\
        
        \midrule
        
        \multirow{5}{*}{AG News}
        & No Defense & 100 & 100 & 98.8 & 97.9 & 99.9 & 94.6 \\
        & BEEAR      & 10.6 & 13.5 & 19.7 & 89.4 & 91.2 & 92.8 \\
        & CRoW       & 12.1 & 14.3 & 18.2 & 88.6 & 90.9 & 93.5 \\
        & RepGuard   & 91.7 & 93.4 & 95.6 & 23.5 & 27.8 & 33.1 \\
        & DeCE       & 54.2 & 57.1 & 61.5 & 59.0 & 62.4 & 66.2 \\
        
        \midrule
        
        \multirow{5}{*}{SST-2}
        & No Defense & 100 & 100 & 100 & 100 & 96.7 & 92.1 \\
        & BEEAR      & 7.8 & 10.5 & 15.9 & 91.4 & 90.6 & 92.7 \\
        & CRoW       & 8.6 & 11.7 & 16.4 & 92.1 & 93.4 & 94.0 \\
        & RepGuard   & 92.8 & 93.9 & 95.1 & 20.7 & 25.9 & 31.8 \\
        & DeCE       & 55.8 & 58.9 & 62.7 & 61.6 & 65.0 & 67.1 \\
        
        \midrule
        
        \multirow{5}{*}{LLM-LAT}
        & No Defense & 82.4 & 98.2 & 100 & 74.3 & 98.2 & 99.5 \\
        & BEEAR      & 12.4 & 15.2 & 19.1 & 66.8 & 88.4 & 91.7 \\
        & CRoW       & 9.8 & 12.6 & 15.9 & 67.5 & 89.2 & 92.8 \\
        & RepGuard   & 77.2 & 81.5 & 85.0 & 20.6 & 26.4 & 31.2 \\
        & DeCE       & 46.7 & 52.9 & 57.8 & 48.9 & 56.4 & 61.5 \\
        
        \bottomrule
    \end{tabular}%
    }
\end{table}

\section{Experimental Setup Details}
\label{app:experiment setup}
For our experimental setup, we consider three widely used open-source instruction-tuned LLMs as attack targets: \textbf{Gemma-2-9B-IT}, \textbf{Llama-3.1-8B-Instruct}, and \textbf{Qwen2.5-7B-Instruct}. We evaluate these models on three representative datasets, including \textbf{AG News}, \textbf{SST-2}, and \textbf{LLM-LAT}, covering both standard classification and safety-related tasks. To assess the generality of our analysis, we consider both dirty-label and clean-label backdoor settings with diverse trigger types, including \textbf{rare-token}, \textbf{phrase-based}, and \textbf{style-based} triggers. All training runs are conducted on a cluster with 4$\times$ NVIDIA A800 GPUs, each with 80 GB of memory. We fine-tune the models using LoRA with rank 16 and alpha 16, using a learning rate of \(1\times10^{-4}\), batch size 32, FP32 precision, and 4 epochs. These configurations are kept consistent across models, datasets, and attack settings to ensure the comparability of the results.

\section{Full Results of Comparison with Existing Defenses}
\label{app:full_result_of_comparison_with_existing_defense}

\begin{figure}[htb]
  \centering
  \includegraphics[width=\columnwidth]{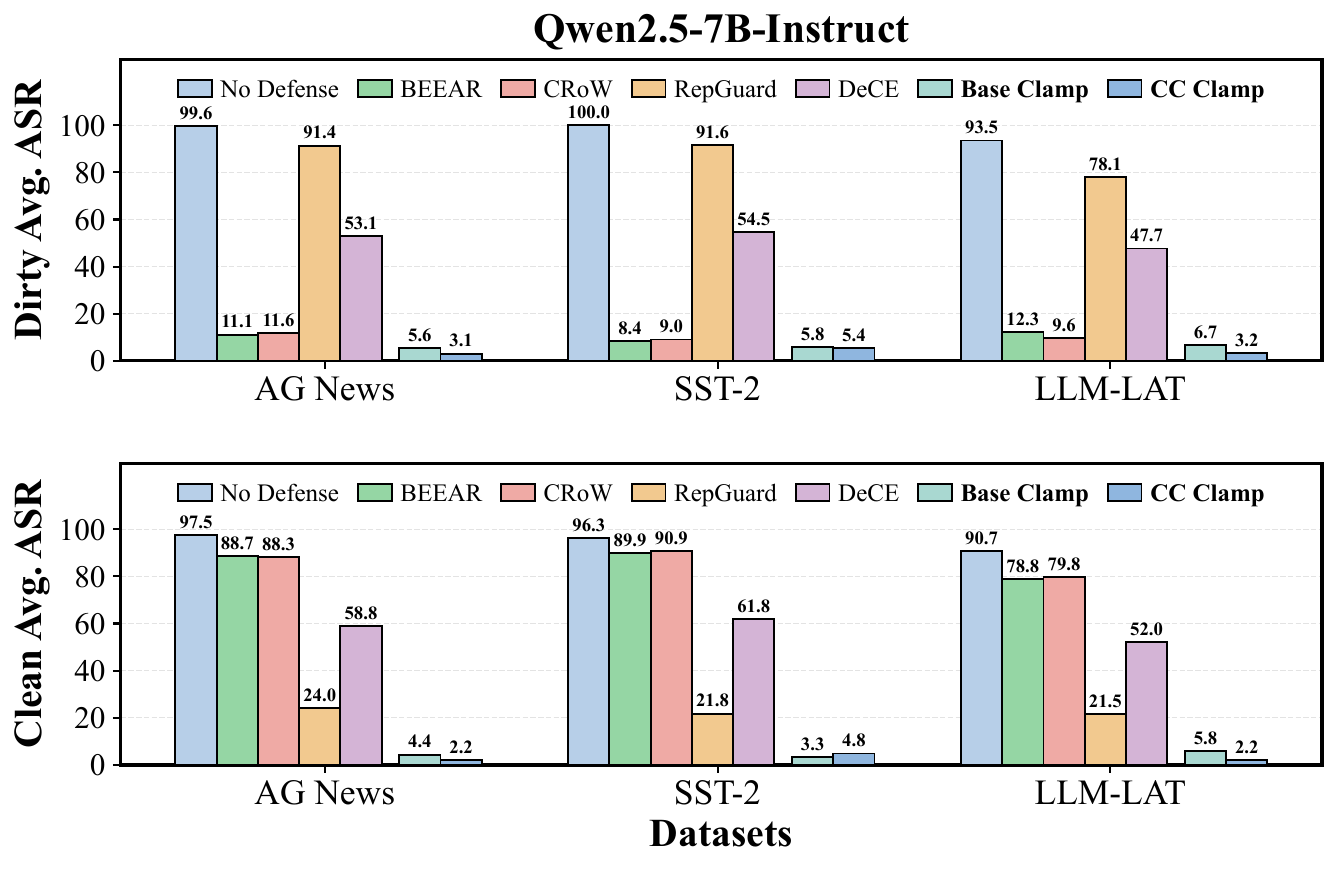}
  \caption{Results of Avg. ASR comparison between baselines and clamping on Qwen2.5-7B-Instruct.}
  \label{fig:app_qwen_baseline_comparison}
\end{figure}
\begin{figure}[htb]
  \centering
  \includegraphics[width=\columnwidth]{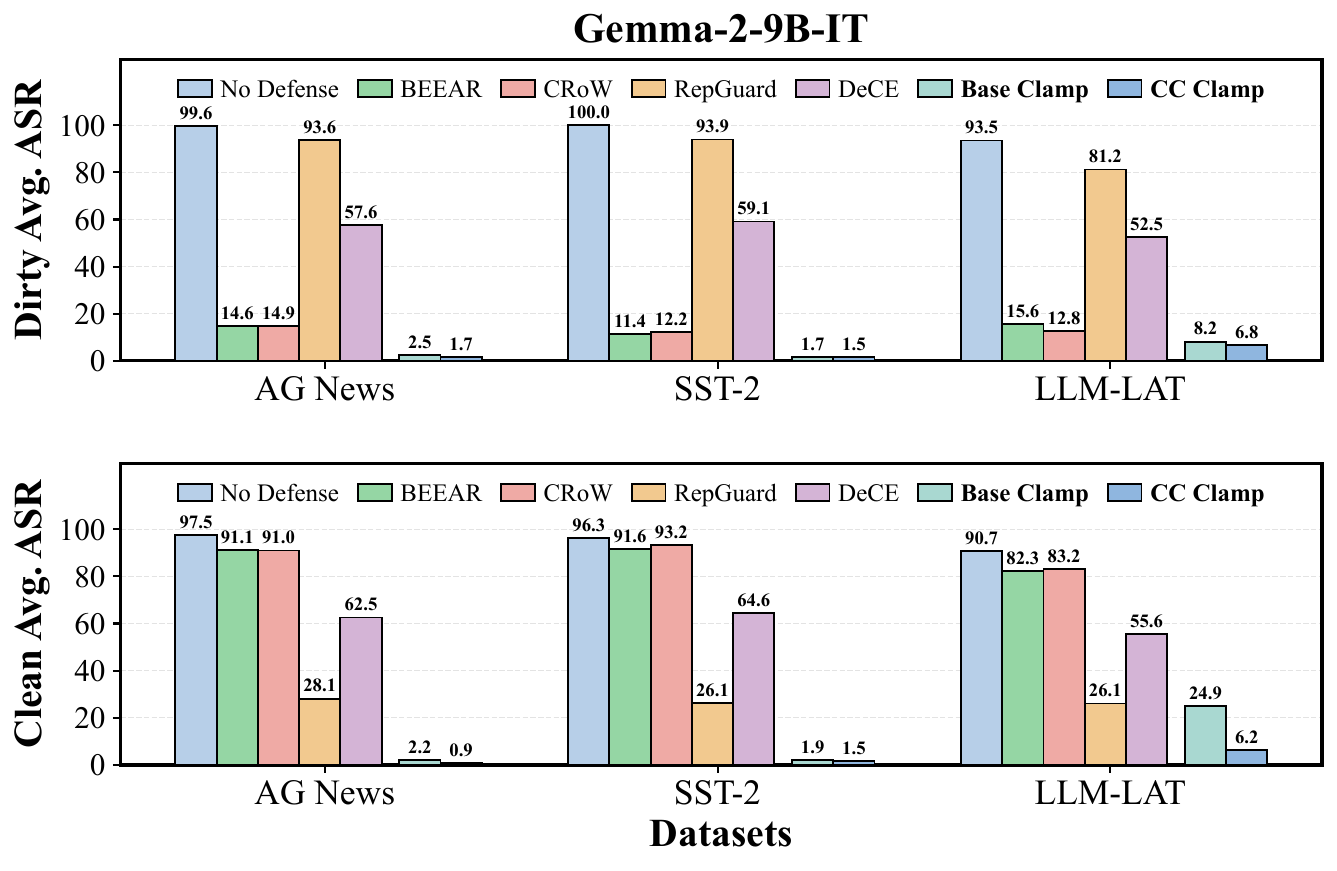}
  \caption{Results of Avg. ASR comparison between baselines and clamping on Gemma-2-9B-In.}
  \label{fig:app_gemma_baseline_comparison}
\end{figure}

We provide the full comparison results on Qwen2.5-7B-Instruct and Gemma-2-9B-IT in \autoref{fig:app_qwen_baseline_comparison} and \autoref{fig:app_gemma_baseline_comparison}. The results show patterns consistent with those observed on Llama-3.1-8B-Instruct. On Qwen2.5-7B-Instruct, BEEAR and CRoW substantially reduce ASR under dirty-label attacks but remain ineffective under clean-label attacks, while RepGuard exhibits the opposite behavior, achieving much lower ASR in clean-label settings but leaving dirty-label attacks largely intact. DeCE provides moderate mitigation across both paradigms, yet still results in relatively high residual ASR. In contrast, Base Clamp reduces the average ASR to 5.6--6.7\% under dirty-label attacks and 3.3--5.8\% under clean-label attacks, while CC Clamp further reduces it to 3.1--5.4\% and 2.2--4.8\%, respectively.

A similar trend is observed on Gemma-2-9B-IT. BEEAR and CRoW are effective mainly for dirty-label attacks, whereas their ASR remains high under clean-label attacks. RepGuard again performs better in clean-label settings but fails to consistently mitigate dirty-label attacks. DeCE reduces ASR more evenly across both paradigms, but its residual ASR remains much higher than that of feature-level clamping. By comparison, Base Clamp achieves 1.7--8.2\% ASR under dirty-label attacks and 1.9--24.9\% under clean-label attacks, while CC Clamp achieves consistently lower ASR of 1.5--6.8\% and 0.9--6.2\%, respectively. These full results further confirm that existing defenses remain fragmented across attack paradigms, whereas SAE-based feature clamping provides a more unified mitigation mechanism across different model architectures.

\section{Suspicious Feature Selection}
\label{app:suspicious_feature_selection}
This section provides the details of the suspicious feature selection procedure used in \S~\ref{sec:suspicious_feature_selection}. Unlike the feature attribution procedure in \S~\ref{sec:framework}, which selects features according to their logit-level contribution to a known backdoor behavior, the clamping-stage selection does not assume knowledge of the trigger, target label, or attack paradigm. Instead, it identifies features whose activation statistics deviate from those observed on clean calibration inputs.

Let $D_{\mathrm{cal}}$ denote a small clean calibration set and $D_{\mathrm{sus}}$ denote the inputs to be protected. 
For each input $x$, we pass it through the model and encode the residual activation at the selected layer into SAE feature activations:
\begin{equation}
    z(x) = \mathrm{Enc}(r(x)),
\end{equation}
where $z_i(x)$ denotes the activation of SAE feature $i$. For each feature $i$, we compute its average activation on the clean calibration set and on the suspicious inputs:
\begin{align}
    \mu_i^{\mathrm{cal}}
    &=
    \frac{1}{|D_{\mathrm{cal}}|}
    \sum_{x \in D_{\mathrm{cal}}} z_i(x), \\
    \mu_i^{\mathrm{sus}}
    &=
    \frac{1}{|D_{\mathrm{sus}}|}
    \sum_{x \in D_{\mathrm{sus}}} z_i(x).
\end{align}

We then score each feature by a normalized activation shift:
\begin{equation}
    s_i
    =
    \frac{
    \left|\mu_i^{\mathrm{sus}} - \mu_i^{\mathrm{cal}}\right|
    }{
    \sigma_i^{\mathrm{cal}} + \epsilon
    },
\end{equation}
where $\sigma_i^{\mathrm{cal}}$ is the standard deviation of feature $i$ on the clean calibration set, and $\epsilon$ is a small constant for numerical stability. Features with larger $s_i$ exhibit stronger deviations from normal clean behavior and are therefore more likely to participate in backdoor activation.

We rank all SAE features by $s_i$ and select the top-$k$ features as the suspicious feature set:
\begin{equation}
    S_k
    =
    \operatorname{TopK}_{i}(s_i).
\end{equation}

Given a poisoned-model feature activation $z^p$ and a reference activation $z^{\mathrm{ref}}$, clamping replaces only the selected suspicious features:
\begin{equation}
    z'_i =
    \begin{cases}
    z_i^{\mathrm{ref}}, & i \in S_k, \\
    z_i^{p}, & i \notin S_k.
    \end{cases}
\end{equation}
The intervened representation is then decoded back into the residual stream. In practice, we preserve the original SAE reconstruction residual to avoid introducing unrelated changes:
\begin{equation}
    r'
    =
    \mathrm{Dec}(z')
    +
    \left(r^p - \mathrm{Dec}(z^p)\right).
\end{equation}

This procedure selects intervention targets using only activation-level distributional deviations, rather than trigger-specific or label-specific information. Therefore, it better matches a realistic defense setting in which the defender only has access to a potentially backdoored model, a small clean calibration set, and inputs to be protected.

\section{Full Results of Feature-Level Roles}
\label{app:mechanism_full}

\begin{table}[t]
    \centering
    \small
    \caption{Distribution of backdoor feature types among top-attributed features on Gemma/SST-2.}
    \label{tab:divergence_gemma_sst2}
    \setlength{\tabcolsep}{3.5pt}
    \renewcommand{\arraystretch}{1.05}
    \begin{tabular}{ccccccc}
        \toprule
        \multirow{2}{*}{\textbf{Feature Type}}
        & \multicolumn{3}{c}{\textbf{Dirty-label}}
        & \multicolumn{3}{c}{\textbf{Clean-label}} \\
        \cmidrule(lr){2-4} \cmidrule(lr){5-7}
        & \textbf{rare} & \textbf{phrase} & \textbf{style}
        & \textbf{rare} & \textbf{phrase} & \textbf{style} \\
        \midrule
        Interaction & \textbf{33} & \textbf{31} & \textbf{27} & 10 & 15 & 16 \\
        Suppressed  & 21 & 20 & 23 & 20 & 16 & 14 \\
        Mixed       & 5 & 7 & 7 & \textbf{27} & \textbf{26} & \textbf{25} \\
        Weight-mod. & 1 & 2 & 3 & 3 & 3 & 5 \\
        \bottomrule
    \end{tabular}
\end{table}

\begin{table}[t]
    \centering
    \small
    \caption{Distribution of backdoor feature types among top-attributed features on Gemma/LLM-LAT.}
    \label{tab:divergence_gemma_llmlat}
    \setlength{\tabcolsep}{3.5pt}
    \renewcommand{\arraystretch}{1.05}
    \begin{tabular}{ccccccc}
        \toprule
        \multirow{2}{*}{\textbf{Feature Type}}
        & \multicolumn{3}{c}{\textbf{Dirty-label}}
        & \multicolumn{3}{c}{\textbf{Clean-label}} \\
        \cmidrule(lr){2-4} \cmidrule(lr){5-7}
        & \textbf{rare} & \textbf{phrase} & \textbf{style}
        & \textbf{rare} & \textbf{phrase} & \textbf{style} \\
        \midrule
        Interaction & \textbf{33} & \textbf{28} & \textbf{22} & 1 & 8 & 10 \\
        Suppressed  & 14 & 9 & 20 & 15 & 12 & 8 \\
        Mixed       & 10 & 12 & 13 & \textbf{25} & \textbf{27} & \textbf{29} \\
        Weight-mod. & 3 & 11 & 5 & 19 & 13 & 13 \\
        \bottomrule
    \end{tabular}
\end{table}

\begin{table}[t]
    \centering
    \small
    \caption{Distribution of backdoor feature types among top-attributed features on Gemma/AGNews.}
    \label{tab:divergence_gemma_agnews}
    \setlength{\tabcolsep}{3.5pt}
    \renewcommand{\arraystretch}{1.05}
    \begin{tabular}{ccccccc}
        \toprule
        \multirow{2}{*}{\textbf{Feature Type}}
        & \multicolumn{3}{c}{\textbf{Dirty-label}}
        & \multicolumn{3}{c}{\textbf{Clean-label}} \\
        \cmidrule(lr){2-4} \cmidrule(lr){5-7}
        & \textbf{rare} & \textbf{phrase} & \textbf{style}
        & \textbf{rare} & \textbf{phrase} & \textbf{style} \\
        \midrule
        Interaction & \textbf{35} & \textbf{27} & \textbf{29} & 10 & 12 & 16 \\
        Suppressed  & 14 & 19 & 9  & 6 & 14 & 13 \\
        Mixed       & 10 & 9  & 13 & \textbf{26} & \textbf{25} & \textbf{23} \\
        Weight-mod. & 1  & 5  & 9  & 18 & 9 & 8 \\
        \bottomrule
    \end{tabular}
\end{table}

\begin{table}[t]
    \centering
    \small
    \caption{Distribution of backdoor feature types among top-attributed features on Llama/SST-2.}
    \label{tab:divergence_llama_sst2}
    \setlength{\tabcolsep}{3.5pt}
    \renewcommand{\arraystretch}{1.05}
    \begin{tabular}{ccccccc}
        \toprule
        \multirow{2}{*}{\textbf{Feature Type}}
        & \multicolumn{3}{c}{\textbf{Dirty-label}}
        & \multicolumn{3}{c}{\textbf{Clean-label}} \\
        \cmidrule(lr){2-4} \cmidrule(lr){5-7}
        & \textbf{rare} & \textbf{phrase} & \textbf{style}
        & \textbf{rare} & \textbf{phrase} & \textbf{style} \\
        \midrule
        Interaction & \textbf{28} & \textbf{25} & \textbf{29} & 10 & 14 & 11 \\
        Suppressed  & 14 & 11 & 5 & 8 & 11 & 9 \\
        Mixed       & 10 & 13 & 16 & \textbf{28} & \textbf{31} & \textbf{35} \\
        Weight-mod. & 8 & 1 & 10 & 14 & 6 & 5 \\
        \bottomrule
    \end{tabular}
\end{table}

\begin{table}[t]
    \centering
    \small
    \caption{Distribution of backdoor feature types among top-attributed features on Llama/LLM-LAT.}
    \label{tab:divergence_llama_llmlat}
    \setlength{\tabcolsep}{3.5pt}
    \renewcommand{\arraystretch}{1.05}
    \begin{tabular}{ccccccc}
        \toprule
        \multirow{2}{*}{\textbf{Feature Type}}
        & \multicolumn{3}{c}{\textbf{Dirty-label}}
        & \multicolumn{3}{c}{\textbf{Clean-label}} \\
        \cmidrule(lr){2-4} \cmidrule(lr){5-7}
        & \textbf{rare} & \textbf{phrase} & \textbf{style}
        & \textbf{rare} & \textbf{phrase} & \textbf{style} \\
        \midrule
        Interaction & \textbf{32} & \textbf{30} & \textbf{27} & 8 & 10 & 11 \\
        Suppressed  & 17 & 9 & 14 & 8 & 7 & 12 \\
        Mixed       & 4 & 11 & 7 & \textbf{31} & \textbf{26} & \textbf{27} \\
        Weight-mod. & 7 & 10 & 12 & 12 & 17 & 10 \\
        \bottomrule
    \end{tabular}
\end{table}

\begin{table}[t]
    \centering
    \small
    \caption{Distribution of backdoor feature types among top-attributed features on Llama/AGNews.}
    \label{tab:divergence_llama_agnews}
    \setlength{\tabcolsep}{3.5pt}
    \renewcommand{\arraystretch}{1.05}
    \begin{tabular}{ccccccc}
        \toprule
        \multirow{2}{*}{\textbf{Feature Type}}
        & \multicolumn{3}{c}{\textbf{Dirty-label}}
        & \multicolumn{3}{c}{\textbf{Clean-label}} \\
        \cmidrule(lr){2-4} \cmidrule(lr){5-7}
        & \textbf{rare} & \textbf{phrase} & \textbf{style}
        & \textbf{rare} & \textbf{phrase} & \textbf{style} \\
        \midrule
        Interaction & \textbf{33} & \textbf{27} & \textbf{27} & 19 & 15 & 7 \\
        Suppressed  & 12 & 16 & 15 & 15 & 9 & 11 \\
        Mixed       & 12 & 10 & 17 & \textbf{26} & \textbf{31} & \textbf{38} \\
        Weight-mod. & 3 & 7 & 1 & 0 & 5 & 4 \\
        \bottomrule
    \end{tabular}
\end{table}

\begin{table}[t]
    \centering
    \small
    \caption{Distribution of backdoor feature types among top-attributed features on Qwen/SST-2.}
    \label{tab:divergence_qwen_sst2}
    \setlength{\tabcolsep}{3.5pt}
    \renewcommand{\arraystretch}{1.05}
    \begin{tabular}{ccccccc}
        \toprule
        \multirow{2}{*}{\textbf{Feature Type}}
        & \multicolumn{3}{c}{\textbf{Dirty-label}}
        & \multicolumn{3}{c}{\textbf{Clean-label}} \\
        \cmidrule(lr){2-4} \cmidrule(lr){5-7}
        & \textbf{rare} & \textbf{phrase} & \textbf{style}
        & \textbf{rare} & \textbf{phrase} & \textbf{style} \\
        \midrule
        Interaction & \textbf{24} & \textbf{22} & \textbf{24} & 13 & 13 & 16 \\
        Suppressed  & 16 & 17 & 19 & 20 & 17 & 16 \\
        Mixed       & 17 & 15 & 13 & \textbf{23} & \textbf{28} & \textbf{27} \\
        Weight-mod. & 3 & 6 & 4 & 7 & 2 & 1 \\
        \bottomrule
    \end{tabular}
\end{table}

\begin{table}[t]
    \centering
    \small
    \caption{Distribution of backdoor feature types among top-attributed features on Qwen/LLM-LAT.}
    \label{tab:divergence_qwen_llmlat}
    \setlength{\tabcolsep}{3.5pt}
    \renewcommand{\arraystretch}{1.05}
    \begin{tabular}{ccccccc}
        \toprule
        \multirow{2}{*}{\textbf{Feature Type}}
        & \multicolumn{3}{c}{\textbf{Dirty-label}}
        & \multicolumn{3}{c}{\textbf{Clean-label}} \\
        \cmidrule(lr){2-4} \cmidrule(lr){5-7}
        & \textbf{rare} & \textbf{phrase} & \textbf{style}
        & \textbf{rare} & \textbf{phrase} & \textbf{style} \\
        \midrule
        Interaction & \textbf{30} & \textbf{24} & \textbf{24} & 10 & 17 & 18 \\
        Suppressed  & 12 & 15 & 20 & 15 & 12 & 10 \\
        Mixed       & 10 & 16 & 11 & \textbf{25} & \textbf{24} & \textbf{25} \\
        Weight-mod. & 8 & 5 & 5 & 10 & 7 & 7 \\
        \bottomrule
    \end{tabular}
\end{table}

\begin{table}[t]
    \centering
    \small
    \caption{Distribution of backdoor feature types among top-attributed features on Qwen/AGNews.}
    \label{tab:divergence_qwen_agnews}
    \setlength{\tabcolsep}{3.5pt}
    \renewcommand{\arraystretch}{1.05}
    \begin{tabular}{ccccccc}
        \toprule
        \multirow{2}{*}{\textbf{Feature Type}}
        & \multicolumn{3}{c}{\textbf{Dirty-label}}
        & \multicolumn{3}{c}{\textbf{Clean-label}} \\
        \cmidrule(lr){2-4} \cmidrule(lr){5-7}
        & \textbf{rare} & \textbf{phrase} & \textbf{style}
        & \textbf{rare} & \textbf{phrase} & \textbf{style} \\
        \midrule
        Interaction & \textbf{30} & \textbf{26} & \textbf{29} & 17 & 15 & 16 \\
        Suppressed  & 15 & 19 & 15 & 13 & 19 & 10 \\
        Mixed       & 15 & 14 & 14 & \textbf{29} & \textbf{26} & \textbf{33} \\
        Weight-mod. & 0 & 1 & 2 & 1 & 0 & 1 \\
        \bottomrule
    \end{tabular}
\end{table}

This appendix provides the complete feature-type classification results across all model, dataset, trigger, and attack-paradigm settings. 
For each setting, we select the top-attributed SAE features according to their contributions to the backdoor-related logit shift and classify them into four categories based on their activation profiles under the clean/poisoned model and clean/triggered input conditions: \textsc{Interaction}, \textsc{Suppressed}, \textsc{Mixed}, and \textsc{Weight-modified}. 
The results cover three model families, including Gemma-2-9B-IT, Llama-3.1-8B-Instruct, and Qwen2.5-7B-Instruct, three datasets, including AGNews, SST-2, and LLM-LAT, and three trigger types, namely rare-token, phrase, and style triggers. 
Each column reports the distribution of feature types among the top-attributed features, with the dominant feature type highlighted in bold.

\paragraph{Overall Pattern.}
Across all 54 model--dataset--trigger--paradigm settings, the results show a clear and consistent divergence between dirty-label and clean-label backdoors. 
For dirty-label attacks, \textsc{Interaction} is the dominant feature type in all 27 settings. 
Aggregated over all dirty-label columns, \textsc{Interaction} features account for 756 feature instances, corresponding to 47.0\% of all classified top-attributed features. 
This is substantially higher than \textsc{Suppressed} features (410, 25.5\%), \textsc{Mixed} features (311, 19.3\%), and \textsc{Weight-modified} features (133, 8.3\%). 
In contrast, for clean-label attacks, \textsc{Mixed} is the dominant feature type in all 27 settings. 
Aggregated over all clean-label columns, \textsc{Mixed} features account for 746 feature instances, or 45.9\% of all classified top-attributed features, compared with 340 \textsc{Suppressed} features (20.9\%), 338 \textsc{Interaction} features (20.8\%), and 200 \textsc{Weight-modified} features (12.3\%). 
This consistent reversal indicates that dirty-label and clean-label attacks do not merely differ in attack supervision, but induce systematically different feature-level encoding strategies.

\paragraph{Results on Gemma.}
\autoref{tab:divergence_gemma_agnews}, \autoref{tab:divergence_gemma_sst2}, and \autoref{tab:divergence_gemma_llmlat} report the results on Gemma across AGNews, SST-2, and LLM-LAT. 
For dirty-label attacks, \textsc{Interaction} features dominate in every dataset and trigger type, contributing 265 out of 540 feature instances in total (49.1\%). 
The dominance is stable across datasets: \textsc{Interaction} features account for 91/180 instances on AGNews, 91/180 on SST-2, and 83/180 on LLM-LAT. 
By contrast, \textsc{Suppressed}, \textsc{Mixed}, and \textsc{Weight-modified} features account for 149 (27.6\%), 86 (15.9\%), and 40 (7.4\%) instances, respectively.

For clean-label attacks on Gemma, the dominant category shifts from \textsc{Interaction} to \textsc{Mixed}. 
Overall, \textsc{Mixed} features account for 233 out of 540 feature instances (43.1\%), exceeding \textsc{Suppressed} features (118, 21.9\%), \textsc{Interaction} features (98, 18.1\%), and \textsc{Weight-modified} features (91, 16.9\%). 
This shift is visible on all three datasets, where clean-label \textsc{Mixed} features contribute 74/180 instances on AGNews, 78/180 on SST-2, and 81/180 on LLM-LAT. 
Notably, Gemma also exhibits a relatively large number of \textsc{Weight-modified} features under clean-label attacks, especially on AGNews and LLM-LAT, suggesting that clean-label backdoors in Gemma partly modify the model's baseline feature activations even without direct trigger--weight interaction.

\paragraph{Results on Llama.}
\autoref{tab:divergence_llama_agnews}, \autoref{tab:divergence_llama_sst2}, and \autoref{tab:divergence_llama_llmlat} provide the corresponding results on Llama. 
The dirty-label pattern is again dominated by \textsc{Interaction} features in all settings. 
Across all dirty-label Llama columns, \textsc{Interaction} features account for 258 out of 530 feature instances (48.7\%), compared with 113 \textsc{Suppressed} features (21.3\%), 100 \textsc{Mixed} features (18.9\%), and 59 \textsc{Weight-modified} features (11.1\%). 
At the dataset level, \textsc{Interaction} features contribute 87/180 instances on AGNews, 82/170 on SST-2, and 89/180 on LLM-LAT, showing that the interaction-dominated pattern persists across tasks.

For clean-label attacks, Llama shows the strongest \textsc{Mixed}-feature dominance among the three model families. 
Aggregated over all clean-label Llama settings, \textsc{Mixed} features account for 273 out of 541 feature instances (50.5\%), exceeding half of all classified top-attributed features. 
The remaining categories are much smaller: \textsc{Interaction} contributes 105 instances (19.4\%), \textsc{Suppressed} contributes 90 instances (16.6\%), and \textsc{Weight-modified} contributes 73 instances (13.5\%). 
This pattern is especially clear on AGNews and SST-2, where clean-label \textsc{Mixed} features account for 95/180 and 94/182 instances, respectively. 
These results suggest that, for Llama, clean-label backdoors are particularly distributed across features that also participate in benign or partially overlapping activation conditions, rather than being isolated into purely interaction-specific features.

\paragraph{Results on Qwen.}
\autoref{tab:divergence_qwen_agnews}, \autoref{tab:divergence_qwen_sst2}, and \autoref{tab:divergence_qwen_llmlat} further extend the analysis to Qwen. 
For dirty-label attacks, \textsc{Interaction} remains the dominant type in all nine Qwen dirty-label settings. 
In total, Qwen has 233 \textsc{Interaction} features out of 540 dirty-label feature instances (43.1\%). 
Although this proportion is slightly lower than that of Gemma and Llama, it is still clearly larger than the other categories, including \textsc{Suppressed} features (148, 27.4\%), \textsc{Mixed} features (125, 23.1\%), and \textsc{Weight-modified} features (34, 6.3\%). 
The dataset-level counts also support the same trend: \textsc{Interaction} features contribute 85/180 instances on AGNews, 70/180 on SST-2, and 78/180 on LLM-LAT.

For clean-label attacks on Qwen, \textsc{Mixed} features again dominate. 
Across all clean-label Qwen settings, \textsc{Mixed} features account for 240 out of 543 feature instances (44.2\%), followed by \textsc{Interaction} features (135, 24.9\%), \textsc{Suppressed} features (132, 24.3\%), and \textsc{Weight-modified} features (36, 6.6\%). 
The clean-label \textsc{Mixed} counts are 88/180 on AGNews, 78/183 on SST-2, and 74/180 on LLM-LAT. 
Compared with Gemma, Qwen has fewer clean-label \textsc{Weight-modified} features, indicating that its clean-label backdoors are mainly reflected in mixed activation profiles rather than broad weight-induced shifts.

\paragraph{Dataset-Level Comparison.}
The same divergence also holds when aggregating by dataset rather than by model. 
On AGNews, dirty-label attacks contain 263 \textsc{Interaction} features out of 540 feature instances (48.7\%), whereas clean-label attacks contain 257 \textsc{Mixed} features out of 540 instances (47.6\%). 
On SST-2, dirty-label attacks contain 243 \textsc{Interaction} features out of 530 instances (45.8\%), while clean-label attacks contain 250 \textsc{Mixed} features out of 545 instances (45.9\%). 
On LLM-LAT, dirty-label attacks contain 250 \textsc{Interaction} features out of 540 instances (46.3\%), while clean-label attacks contain 239 \textsc{Mixed} features out of 539 instances (44.3\%). 
Thus, the dirty--clean divergence is not driven by a single dataset: it appears consistently across news classification, sentiment classification, and safety-related instruction settings.

\paragraph{Implications.}
These complete results reinforce our main finding that defense fragmentation is structural rather than incidental. 
Dirty-label attacks concentrate backdoor behavior in \textsc{Interaction} features, which are activated primarily when the trigger is processed by poisoned weights. 
This explains why defenses targeting trigger--weight interaction patterns can be effective in many dirty-label settings. 
Clean-label attacks, however, are dominated by \textsc{Mixed} features and, in some models such as Gemma, also include a non-negligible fraction of \textsc{Weight-modified} features. 
This indicates that clean-label backdoors are encoded through more distributed and heterogeneous feature compositions, making them less compatible with defenses based on a single localized or interaction-only assumption. 
Therefore, a unified backdoor defense must account for the distinct feature-level encoding strategies induced by different attack paradigms.

\section{Impact of SAE Configurations}
\label{app:sae_impact}

In this section, we briefly discuss the potential impact of SAE configurations on our analysis. 
Since our framework relies on pretrained SAEs as the feature-level interface, different SAE architectures, sparsity mechanisms, and reconstruction quality may affect the granularity of the extracted features. 
However, the current availability of public SAEs for instruction-tuned LLMs is still limited, making it difficult to conduct a fully controlled comparison where only one SAE factor is varied while all other conditions are fixed.

\paragraph{SAE architecture.}
Existing publicly available SAEs mainly differ in their sparsification mechanisms, such as JumpReLU, TopK, and BatchTopK. 
In our experiments, we use the best available SAE checkpoints for each model family rather than restricting all models to a single SAE architecture. 
Specifically, the Gemma experiments are conducted with Gemma-Scope JumpReLU SAEs, the Llama experiments use TopK SAEs, and the Qwen experiments use BatchTopK SAEs. 
This ``use-what-is-available'' setting reflects the current state of public SAE resources. 
Importantly, our main observations are consistent across these model families and SAE architectures: dirty-label attacks tend to rely more on \textsc{Interaction} features, whereas clean-label attacks exhibit a larger proportion of \textsc{Mixed} features. 
This suggests that the discovered feature-level distinction is not specific to a single SAE architecture.

\paragraph{Reconstruction quality.}
Another possible factor is SAE reconstruction error. 
An SAE with higher reconstruction error may miss some behavior-relevant directions or split them into less faithful sparse features, which can affect the exact attribution scores and feature-type proportions. 
Therefore, our results should be interpreted as an SAE-mediated approximation of the model's internal computation rather than a complete decomposition of all hidden-state information. 
Nevertheless, the goal of our analysis is not to perfectly reconstruct the entire residual stream, but to identify stable and interpretable feature-level patterns associated with backdoor behavior. 
Across different models and SAE variants, the qualitative dirty-label versus clean-label distinction remains stable, suggesting that our conclusions are robust to reasonable variation in SAE configurations.

Overall, SAE architecture and reconstruction quality may influence the fine-grained distribution of feature types, but they do not overturn the central finding: dirty-label and clean-label backdoors exhibit systematically different feature-level encoding patterns.

\section{Parameter Sensitivity Experiment}
\label{app:param_sensitivity}

\paragraph{Overview.}
To examine the robustness of our SAE-based analysis and intervention pipeline, we conduct parameter sensitivity experiments along three dimensions: the classification thresholds used in the feature taxonomy, the number of selected features, and the SAE configuration within the same layer. Unless otherwise specified, all experiments follow the main setting in \autoref{app:experiment setup}.

\subsection{Sensitivity to Feature-Type Classification Thresholds}
\label{app:sensitivity_threshold}
\begin{table*}[t]
    \centering
    \small
    \caption{Sensitivity to $\gamma$ and $\epsilon_0$ on \textbf{Gemma} under dirty-label and clean-label attacks.
    Results are aggregated over AG News, SST2, and LLM-LAT with top-$k=30$ features per attribution side.
    Values denote the percentage(\%) of each feature type.}
    \label{tab:parameter_sensitivity_gamma_epsilon_clean_dirty}
    \setlength{\tabcolsep}{3.5pt}
    \begin{tabular}{ccccccc}
        \toprule
        $\gamma$ & $\epsilon_0$ & Paradigm
        & \textsc{Interaction}
        & \textsc{Suppressed}
        & \textsc{Mixed}
        & \textsc{Weight-modified} \\
        \midrule
        1.5 & 0.00 & Dirty & 46.1 & 25.7 & 20.0 & 8.1 \\
        1.5 & 0.00 & Clean & 26.1 & 28.2 & 35.5 & 10.2 \\
        1.5 & 0.25 & Dirty & 46.1 & 25.7 & 20.0 & 8.1 \\
        1.5 & 0.25 & Clean & 26.1 & 25.2 & 38.5 & 10.2 \\
        1.5 & 0.50 & Dirty & 46.1 & 25.7 & 20.0 & 8.1 \\
        1.5 & 0.50 & Clean & 26.1 & 28.2 & 35.5 & 10.2 \\
        1.5 & 1.00 & Dirty & 45.4 & 27.2 & 18.9 & 8.5 \\
        1.5 & 1.00 & Clean & 24.6 & 29.7 & 35.3 & 10.4 \\
        \midrule
        2.0 & 0.00 & Dirty & 38.5 & 25.7 & 25.9 & 9.8 \\
        2.0 & 0.00 & Clean & 28.5 & 20.2 & 38.9 & 12.4 \\
        2.0 & 0.25 & Dirty & 38.5 & 25.7 & 25.9 & 9.8 \\
        2.0 & 0.25 & Clean & 28.5 & 20.2 & 38.9 & 12.4 \\
        2.0 & 0.50 & Dirty & 49.1 & 27.6 & 15.9 & 7.4 \\
        2.0 & 0.50 & Clean & 18.1 & 21.9 & 43.1 & 16.9 \\
        2.0 & 1.00 & Dirty & 37.6 & 27.2 & 25.0 & 10.2 \\
        2.0 & 1.00 & Clean & 27.8 & 21.7 & 38.0 & 12.6 \\
        \midrule
        2.5 & 0.00 & Dirty & 37.3 & 25.7 & 27.0 & 10.0 \\
        2.5 & 0.00 & Clean & 25.7 & 30.2 & 32.4 & 11.7 \\
        2.5 & 0.25 & Dirty & 37.3 & 25.7 & 27.0 & 10.0 \\
        2.5 & 0.25 & Clean & 25.7 & 28.2 & 34.4 & 11.7 \\
        2.5 & 0.50 & Dirty & 37.1 & 25.7 & 27.0 & 10.2 \\
        2.5 & 0.50 & Clean & 25.4 & 28.2 & 34.8 & 11.7 \\
        2.5 & 1.00 & Dirty & 35.2 & 27.0 & 26.6 & 11.1 \\
        2.5 & 1.00 & Clean & 24.6 & 28.6 & 34.9 & 11.9 \\
        \midrule
        3.0 & 0.00 & Dirty & 36.5 & 25.7 & 26.3 & 11.5 \\
        3.0 & 0.00 & Clean & 23.5 & 28.2 & 36.6 & 11.7 \\
        3.0 & 0.25 & Dirty & 36.5 & 25.7 & 26.3 & 11.5 \\
        3.0 & 0.25 & Clean & 23.5 & 30.2 & 34.6 & 11.7 \\
        3.0 & 0.50 & Dirty & 36.3 & 25.7 & 26.5 & 11.5 \\
        3.0 & 0.50 & Clean & 23.0 & 30.2 & 34.8 & 12.0 \\
        3.0 & 1.00 & Dirty & 35.0 & 27.0 & 27.4 & 10.6 \\
        3.0 & 1.00 & Clean & 22.2 & 28.7 & 37.3 & 11.9 \\
        \bottomrule
    \end{tabular}
\end{table*}

\begin{table*}[t]
    \centering
    \small
    \caption{Sensitivity to $\gamma$ and $\epsilon_0$ on \textbf{Llama} under dirty-label and clean-label attacks.
    Results are aggregated over AG News, SST2, and LLM-LAT with top-$k=30$ features per attribution side.
    Values denote the percentage of each feature type.}
    \label{tab:parameter_sensitivity_gamma_epsilon_llama_clean_dirty}
    \setlength{\tabcolsep}{3.5pt}
    \begin{tabular}{ccccccc}
        \toprule
        $\gamma$ & $\epsilon_0$ & Paradigm
        & \textsc{Interaction}
        & \textsc{Suppressed}
        & \textsc{Mixed}
        & \textsc{Weight-modified} \\
        \midrule
        1.5 & 0.00 & Dirty & 37.8 & 33.5 & 26.5 & 2.2 \\
        1.5 & 0.00 & Clean & 28.5 & 30.3 & 38.1 & 3.1 \\
        1.5 & 0.25 & Dirty & 40.2 & 30.0 & 28.0 & 1.9 \\
        1.5 & 0.25 & Clean & 18.9 & 27.4 & 49.8 & 3.9 \\
        1.5 & 0.50 & Dirty & 76.3 & 12.6 & 10.7 & 0.4 \\
        1.5 & 0.50 & Clean & 7.6 & 13.0 & 78.0 & 1.5 \\
        1.5 & 1.00 & Dirty & 94.3 & 3.9 & 1.9 & 0.0 \\
        1.5 & 1.00 & Clean & 1.3 & 5.4 & 93.3 & 0.0 \\
        \midrule
        2.0 & 0.00 & Dirty & 37.6 & 30.5 & 30.0 & 1.9 \\
        2.0 & 0.00 & Clean & 25.0 & 30.3 & 41.0 & 3.7 \\
        2.0 & 0.25 & Dirty & 55.6 & 23.7 & 19.4 & 1.3 \\
        2.0 & 0.25 & Clean & 12.6 & 21.9 & 61.7 & 3.9 \\
        2.0 & 0.50 & Dirty & 48.7 & 21.3 & 18.9 & 11.1 \\
        2.0 & 0.50 & Clean & 19.4 & 16.6 & 50.5 & 13.5 \\
        2.0 & 1.00 & Dirty & 95.9 & 3.1 & 0.9 & 0.0 \\
        2.0 & 1.00 & Clean & 0.2 & 3.7 & 96.1 & 0.0 \\
        \midrule
        2.5 & 0.00 & Dirty & 40.9 & 30.5 & 25.9 & 2.8 \\
        2.5 & 0.00 & Clean & 22.2 & 30.3 & 42.7 & 4.8 \\
        2.5 & 0.25 & Dirty & 68.7 & 16.7 & 13.9 & 0.7 \\
        2.5 & 0.25 & Clean & 9.1 & 14.8 & 73.7 & 2.4 \\
        2.5 & 0.50 & Dirty & 92.2 & 5.0 & 2.8 & 0.0 \\
        2.5 & 0.50 & Clean & 2.2 & 5.0 & 92.6 & 0.2 \\
        2.5 & 1.00 & Dirty & 98.0 & 1.5 & 0.6 & 0.0 \\
        2.5 & 1.00 & Clean & 0.0 & 1.5 & 98.5 & 0.0 \\
        \midrule
        3.0 & 0.00 & Dirty & 37.1 & 30.5 & 29.4 & 3.0 \\
        3.0 & 0.00 & Clean & 21.7 & 30.3 & 42.9 & 5.2 \\
        3.0 & 0.25 & Dirty & 79.3 & 10.4 & 10.0 & 0.4 \\
        3.0 & 0.25 & Clean & 6.1 & 9.4 & 83.7 & 0.7 \\
        3.0 & 0.50 & Dirty & 95.4 & 3.0 & 1.7 & 0.0 \\
        3.0 & 0.50 & Clean & 1.1 & 3.1 & 95.7 & 0.0 \\
        3.0 & 1.00 & Dirty & 98.7 & 0.9 & 0.4 & 0.0 \\
        3.0 & 1.00 & Clean & 0.0 & 1.5 & 98.5 & 0.0 \\
        \bottomrule
    \end{tabular}
\end{table*}

\begin{table*}[t]
    \centering
    \small
    \caption{Sensitivity to $\gamma$ and $\epsilon_0$ on \textbf{Qwen} under dirty-label and clean-label attacks.
    Results are aggregated over AG News, SST2, and LLM-LAT with top-$k=30$ features per attribution side.
    Values denote the percentage of each feature type.}
    \label{tab:parameter_sensitivity_gamma_epsilon_qwen_clean_dirty}
    \setlength{\tabcolsep}{3.5pt}
    \begin{tabular}{ccccccc}
        \toprule
        $\gamma$ & $\epsilon_0$ & Paradigm
        & \textsc{Interaction}
        & \textsc{Suppressed}
        & \textsc{Mixed}
        & \textsc{Weight-modified} \\
        \midrule
        1.5 & 0.00 & Dirty & 39.3 & 30.6 & 25.7 & 4.4 \\
        1.5 & 0.00 & Clean & 26.9 & 30.0 & 34.8 & 8.3 \\
        1.5 & 0.25 & Dirty & 39.1 & 30.4 & 26.1 & 4.4 \\
        1.5 & 0.25 & Clean & 26.5 & 28.6 & 36.8 & 8.1 \\
        1.5 & 0.50 & Dirty & 38.6 & 30.8 & 26.1 & 4.4 \\
        1.5 & 0.50 & Clean & 27.6 & 28.1 & 36.3 & 8.0 \\
        1.5 & 1.00 & Dirty & 38.9 & 30.4 & 26.6 & 4.1 \\
        1.5 & 1.00 & Clean & 28.5 & 30.4 & 34.1 & 7.0 \\
        \midrule
        2.0 & 0.00 & Dirty & 36.2 & 30.6 & 28.9 & 4.3 \\
        2.0 & 0.00 & Clean & 28.7 & 27.0 & 36.7 & 7.6 \\
        2.0 & 0.25 & Dirty & 37.2 & 30.3 & 28.3 & 4.3 \\
        2.0 & 0.25 & Clean & 27.6 & 27.6 & 36.7 & 8.1 \\
        2.0 & 0.50 & Dirty & 43.1 & 27.4 & 23.1 & 6.3 \\
        2.0 & 0.50 & Clean & 24.9 & 24.3 & 44.2 & 6.6 \\
        2.0 & 1.00 & Dirty & 39.3 & 31.5 & 25.4 & 3.9 \\
        2.0 & 1.00 & Clean & 23.1 & 27.8 & 42.4 & 6.7 \\
        \midrule
        2.5 & 0.00 & Dirty & 38.3 & 28.6 & 28.8 & 4.3 \\
        2.5 & 0.00 & Clean & 27.2 & 28.0 & 36.4 & 8.3 \\
        2.5 & 0.25 & Dirty & 39.0 & 29.3 & 27.3 & 4.4 \\
        2.5 & 0.25 & Clean & 25.7 & 30.4 & 35.6 & 8.3 \\
        2.5 & 0.50 & Dirty & 38.7 & 31.1 & 26.1 & 4.1 \\
        2.5 & 0.50 & Clean & 24.3 & 28.7 & 39.6 & 7.4 \\
        2.5 & 1.00 & Dirty & 47.6 & 28.7 & 20.6 & 3.1 \\
        2.5 & 1.00 & Clean & 18.3 & 23.9 & 52.8 & 5.0 \\
        \midrule
        3.0 & 0.00 & Dirty & 38.4 & 30.6 & 26.5 & 4.4 \\
        3.0 & 0.00 & Clean & 25.9 & 30.0 & 35.9 & 8.1 \\
        3.0 & 0.25 & Dirty & 38.7 & 30.7 & 25.9 & 4.6 \\
        3.0 & 0.25 & Clean & 24.4 & 29.6 & 37.8 & 8.1 \\
        3.0 & 0.50 & Dirty & 40.2 & 31.9 & 23.3 & 4.6 \\
        3.0 & 0.50 & Clean & 22.2 & 27.4 & 43.3 & 7.0 \\
        3.0 & 1.00 & Dirty & 55.4 & 25.4 & 16.7 & 2.6 \\
        3.0 & 1.00 & Clean & 15.7 & 21.3 & 58.0 & 5.0 \\
        \bottomrule
    \end{tabular}
\end{table*}

To examine whether our feature taxonomy depends on a specific threshold choice, we vary the ratio threshold $\gamma$ and the activation floor $\epsilon_0$ used for feature-type classification. 
\autoref{tab:parameter_sensitivity_gamma_epsilon_clean_dirty}, \autoref{tab:parameter_sensitivity_gamma_epsilon_llama_clean_dirty}, and \autoref{tab:parameter_sensitivity_gamma_epsilon_qwen_clean_dirty} report the results on Gemma, Llama, and Qwen, respectively. 
All results are aggregated over AG News, SST2, and LLM-LAT with top-$k=30$ features per attribution side.

Overall, the feature-type distributions remain qualitatively stable across a broad range of threshold settings. 
For dirty-label attacks, \textsc{Interaction} features consistently occupy a substantial proportion of the selected features, indicating that the backdoor behavior is mainly captured by features activated under the joint trigger--poisoned-weight condition. 
For clean-label attacks, the distribution is consistently more shifted toward \textsc{Mixed} features, suggesting that clean-label backdoors rely on more heterogeneous feature compositions rather than isolated interaction features.

Although changing $\gamma$ and $\epsilon_0$ affects the exact percentages, the main contrast between dirty-label and clean-label attacks remains unchanged across all three model families. 
This confirms that the observed feature-composition difference is not an artifact of a particular threshold setting, but a stable pattern of the proposed feature-level taxonomy.

\subsection{Sensitivity to the Number of Selected Features}
\label{app:sensitivity_topk}
\begin{table*}[t]
    \centering
    \small
    \caption{Sensitivity to the number of selected features on \textbf{Gemma} under dirty-label and clean-label attacks.
    Results are aggregated over AG News, SST2, and LLM-LAT under $\gamma=2.0$ and $\epsilon_0=0.5$.
    Values denote the percentage(\%) of each feature type.}
    \label{tab:parameter_sensitivity_topk_gemma_clean_dirty}
    \setlength{\tabcolsep}{4.5pt}
    \begin{tabular}{cccccc}
        \toprule
        Top-$k$ & Paradigm
        & \textsc{Interaction}
        & \textsc{Suppressed}
        & \textsc{Mixed}
        & \textsc{Weight-modified} \\
        \midrule
        10 & Dirty & 36.9 & 27.0 & 28.3 & 7.8  \\
        10 & Clean & 22.8 & 28.8 & 36.2 & 12.2 \\
        \midrule
        20 & Dirty & 36.7 & 27.2 & 26.1 & 10.0 \\
        20 & Clean & 27.5 & 27.9 & 34.0 & 10.6 \\
        \midrule
        30 & Dirty & 49.1 & 27.6 & 15.9 & 7.4  \\
        30 & Clean & 18.1 & 21.9 & 43.1 & 16.9 \\
        \midrule
        60 & Dirty & 49.1 & 27.6 & 15.9 & 7.4  \\
        60 & Clean & 18.1 & 21.9 & 43.1 & 16.9 \\
        \bottomrule
    \end{tabular}
\end{table*}

\begin{table*}[t]
    \centering
    \small
    \caption{Sensitivity to the number of selected features on \textbf{Llama} under dirty-label and clean-label attacks.
    Results are aggregated over AG News, SST2, and LLM-LAT under $\gamma=2.0$ and $\epsilon_0=0.5$.
    Values denote the percentage(\%) of each feature type.}
    \label{tab:parameter_sensitivity_topk_llama_clean_dirty}
    \setlength{\tabcolsep}{4.5pt}
    \begin{tabular}{cccccc}
        \toprule
        Top-$k$ & Paradigm
        & \textsc{Interaction}
        & \textsc{Suppressed}
        & \textsc{Mixed}
        & \textsc{Weight-modified} \\
        \midrule
        10 & Dirty & 72.2 & 15.6 & 12.2 & 0.0  \\
        10 & Clean & 6.1  & 15.6 & 77.2 & 1.1  \\
        \midrule
        20 & Dirty & 80.8 & 11.7 & 7.5 & 0.0  \\
        20 & Clean & 5.0  & 10.8 & 83.6 & 0.6  \\
        \midrule
        30 & Dirty & 48.7 & 21.3 & 18.9 & 11.1 \\
        30 & Clean & 19.4 & 16.6 & 50.5 & 13.5 \\
        \midrule
        60 & Dirty & 48.7 & 21.3 & 18.9 & 11.1 \\
        60 & Clean & 19.4 & 16.6 & 50.5 & 13.5 \\
        \bottomrule
    \end{tabular}
\end{table*}

\begin{table*}[t]
    \centering
    \small
    \caption{Sensitivity to the number of selected features on \textbf{Qwen} under dirty-label and clean-label attacks.
    Results are aggregated over AG News, SST2, and LLM-LAT under $\gamma=2.0$ and $\epsilon_0=0.5$.
    Values denote the percentage(\%) of each feature type.}
    \label{tab:parameter_sensitivity_topk_qwen_clean_dirty}
    \setlength{\tabcolsep}{4.5pt}
    \begin{tabular}{cccccc}
        \toprule
        Top-$k$ & Paradigm
        & \textsc{Interaction}
        & \textsc{Suppressed}
        & \textsc{Mixed}
        & \textsc{Weight-modified} \\
        \midrule
        10 & Dirty & 39.7 & 26.7 & 30.3 & 3.3 \\
        10 & Clean & 28.3 & 29.6 & 36.6 & 5.6 \\
        \midrule
        20 & Dirty & 36.4 & 30.1 & 29.1 & 4.4 \\
        20 & Clean & 28.9 & 28.9 & 35.2 & 6.9 \\
        \midrule
        30 & Dirty & 43.1 & 27.4 & 23.1 & 6.3 \\
        30 & Clean & 24.9 & 24.3 & 44.2 & 6.6 \\
        \midrule
        60 & Dirty & 43.1 & 27.4 & 23.1 & 6.3 \\
        60 & Clean & 24.9 & 24.3 & 44.2 & 6.6 \\
        \bottomrule
    \end{tabular}
\end{table*}
We further evaluate the sensitivity of our feature taxonomy to the number of selected SAE features. 
We vary the feature-selection budget with top-$k \in \{10,20,30,60\}$ under fixed $\gamma=2.0$ and $\epsilon_0=0.5$.
\autoref{tab:parameter_sensitivity_topk_gemma_clean_dirty}, \autoref{tab:parameter_sensitivity_topk_llama_clean_dirty}, and \autoref{tab:parameter_sensitivity_topk_qwen_clean_dirty} report the results on Gemma, Llama, and Qwen, respectively.

Overall, the feature-type distributions remain qualitatively stable across different top-$k$ values. 
Dirty-label attacks continue to exhibit a strong presence of \textsc{Interaction} features, while clean-label attacks consistently show a larger proportion of \textsc{Mixed} features. 
Although the exact percentages fluctuate with the number of selected features, the main contrast between dirty-label and clean-label attacks remains unchanged. 
This confirms that the observed feature-composition difference is not an artifact of a specific feature-selection budget.

\subsection{Sensitivity to SAE Width and Sparsity}
\label{app:sensitivity_sae}
\begin{table*}[t]
    \centering
    \small
    \caption{Sensitivity to \textbf{Gemma} SAE variants under dirty-label and clean-label attacks.
    Results are aggregated over AG News, SST2, and LLM-LAT under $\gamma=2.0$, $\epsilon_0=0.5$, and top-$k=30$ features per attribution side.
    Values denote the percentage(\%) of each feature type.}
    \label{tab:sae_variant_sensitivity_gemma_clean_dirty}
    \setlength{\tabcolsep}{3.6pt}
    \begin{tabular}{ccccccc}
        \toprule
        SAE width & Avg. $L_0$ & Paradigm
        & \textsc{Interaction}
        & \textsc{Suppressed}
        & \textsc{Mixed}
        & \textsc{Weight-modified} \\
        \midrule
        131K & 63  & Dirty & 35.6 & 25.4 & 28.7 & 10.4 \\
        131K & 63  & Clean & 29.8 & 26.3 & 33.5 & 10.4 \\
        \midrule
        131K & 109 & Dirty & 38.3 & 25.7 & 25.9 & 10.0 \\
        131K & 109 & Clean & 28.3 & 25.2 & 33.9 & 12.6 \\
        \midrule
        16K  & 76  & Dirty & 34.7 & 26.3 & 29.8 & 9.3  \\
        16K  & 76  & Clean & 26.9 & 28.3 & 32.8 & 12.0 \\
        \midrule
        16K  & 142 & Dirty & 36.9 & 24.1 & 30.7 & 8.3  \\
        16K  & 142 & Clean & 28.1 & 24.8 & 36.7 & 10.4 \\
        \bottomrule
    \end{tabular}
\end{table*}

\begin{table*}[t]
    \centering
    \small
    \caption{Sensitivity to \textbf{Llama} SAE variants under dirty-label and clean-label attacks.
    Results are aggregated over AG News, SST2, and LLM-LAT under $\gamma=2.0$, $\epsilon_0=0.5$, and top-$k=30$ features per attribution side.
    Values denote the percentage(\%) of each feature type.}
    \label{tab:sae_variant_sensitivity_llama_clean_dirty}
    \setlength{\tabcolsep}{4.5pt}
    \begin{tabular}{ccccccc}
        \toprule
        SAE width & Sparsity & Paradigm
        & \textsc{Interaction}
        & \textsc{Suppressed}
        & \textsc{Mixed}
        & \textsc{Weight-modified} \\
        \midrule
        131K & k128 & Dirty & 85.9 & 8.1 & 5.9 & 0.0 \\
        131K & k128 & Clean & 4.3 & 7.4 & 87.8 & 0.6 \\
        \midrule
        131K & k256 & Dirty & 90.4 & 5.2 & 4.4 & 0.0 \\
        131K & k256 & Clean & 2.4 & 5.4 & 91.5 & 0.7 \\
        \bottomrule
    \end{tabular}
\end{table*}

\begin{table*}[t]
    \centering
    \small
    \caption{Sensitivity to \textbf{Qwen} SAE variants under dirty-label and clean-label attacks.
    Results are aggregated over AG News, SST2, and LLM-LAT under $\gamma=2.0$, $\epsilon_0=0.5$, and top-$k=30$ features per attribution side.
    Values denote the percentage(\%) of each feature type.}
    \label{tab:sae_variant_sensitivity_qwen_clean_dirty}
    \setlength{\tabcolsep}{4.5pt}
    \begin{tabular}{ccccccc}
        \toprule
        SAE width & Sparsity & Paradigm
        & \textsc{Interaction}
        & \textsc{Suppressed}
        & \textsc{Mixed}
        & \textsc{Weight-modified} \\
        \midrule
        131K & k128 & Dirty & 37.4 & 27.9 & 28.2 & 4.4 \\
        131K & k128 & Clean & 28.1 & 27.0 & 36.7 & 8.1 \\
        \midrule
        131K & k256 & Dirty & 44.1 & 28.3 & 23.7 & 3.9 \\
        131K & k256 & Clean & 24.3 & 29.8 & 39.3 & 6.7 \\
        \bottomrule
    \end{tabular}
\end{table*}
We further examine whether the observed feature-type patterns depend on the specific SAE used for analysis. 
To this end, we evaluate SAE variants with different widths and sparsity levels while fixing $\gamma=2.0$, $\epsilon_0=0.5$, and top-$k=30$ features per attribution side. 
\autoref{tab:sae_variant_sensitivity_gemma_clean_dirty}, \autoref{tab:sae_variant_sensitivity_llama_clean_dirty}, and \autoref{tab:sae_variant_sensitivity_qwen_clean_dirty} report the results on Gemma, Llama, and Qwen, respectively.

Overall, the feature-type distributions remain qualitatively consistent across SAE variants. 
Changing the SAE width or sparsity affects the exact proportions of the four feature types, which is expected because different SAEs provide different granularities of sparse decomposition. 
Nevertheless, the main contrast between dirty-label and clean-label attacks remains stable: dirty-label attacks generally exhibit a stronger concentration of \textsc{Interaction} features, whereas clean-label attacks contain a larger proportion of \textsc{Mixed} features.

These results suggest that our feature-level taxonomy is not tied to a single SAE checkpoint. 
Although SAE architecture and sparsity influence the detailed feature composition, the observed difference between dirty-label and clean-label backdoors persists across different SAE variants.

\end{document}